\documentclass[12pt,dvips]{article}
\usepackage{amsmath}
\usepackage{epsfig}
\usepackage{amssymb}
\usepackage{color}
\usepackage{amsfonts}
\usepackage{graphicx}%
\providecommand{\U}[1]{\protect\rule{.1in}{.1in}}
\def\lsim{\mathrel{\raise.3ex\hbox{$<$\kern-.75em\lower1ex\hbox{$\sim$}}}}
\def\gsim{\mathrel{\raise.3ex\hbox{$>$\kern-.75em\lower1ex\hbox{$\sim$}}}}

\begin{document}
\begin{titlepage}
\begin{flushright}
hep-ph/yymmnnn\\[1mm]
\today
\end{flushright}
\vskip 1.5cm
\begin{center}
{\Large\bf Measuring superparticle masses at hadron collider using the transverse mass kink }\\[1.5cm]
{Won Sang Cho$^1$, Kiwoon Choi$^1$,Yeong Gyun Kim$^{1,2}$, Chan Beom
Park$^1$}
\end{center}
\vskip 0.5cm
{\small
\begin{center}
$^1$ {\it Department of Physics, KAIST, Daejon 305-701, Korea} \\
$^2$ {\it ARCSEC, Sejong University, Seoul 143-747, Korea}\\
\end{center}
} \vskip 4.cm
\begin{abstract}
\noindent We present a detailed study of the collider observable
$m_{T2}$ applied for pair-produced superparticles decaying to
visible particles and a pair of invisible lightest supersymmetric
particles (LSPs).
Analytic expressions of the maximum of $m_{T2}$ over all events
($m_{T2}^{\rm max}$) are derived.  It is noticed that if the decay
product of each superparticle involves more than one visible
particle, $m_{T2}^{\rm max}$ being a function of the {\it trial}
LSP mass ${m}_\chi$ has a kink structure at ${m}_\chi=$ true LSP
mass, which can be used to determine the mother superparticle mass
and the LSP mass simultaneously. To see how well $m_{T2}^{\rm max}$
can be constructed from collider data,  a Monte-Carlo analysis of
the gluino $m_{T2}$ is performed for some  superparticle spectra.
\end{abstract}
\end{titlepage}


\section{Introduction}
The Large Hadron Collider (LHC) at CERN  will explore soon the TeV
energy scale where new physics beyond the Standard Model (SM) is
likely to reveal itself \cite{atlas,cms}. Among various proposals,
weak scale supersymmetry (SUSY) \cite{susy} is perhaps the most
promising candidate for new physics at TeV as it provides a solution
to the gauge hierarchy problem while complying with gauge coupling
unification. Furthermore, with R-parity conservation, the lightest
supersymmetric particle (LSP) becomes a natural candidate for the
non-baryonic dark matter in the Universe.

Once SUSY signals are discovered through event excess beyond the SM
backgrounds in inclusive search channels, the next step will be the
measurement of SUSY particle masses and other physical properties
through various exclusive decay chains \cite{lhclc}. Then, it might be possible
to reconstruct the underlying SUSY theory, in particular the soft
SUSY breaking terms, using the observed SUSY particle masses. In
this regard, experimental information on gaugino masses can be
particularly useful for distinguishing different SUSY breaking
schemes as the theoretical predictions of low energy gaugino masses
are quite robust compared to those on sfermion masses
\cite{choi-nilles}.

In a recent paper \cite{mt203}, we have examined  the collider
observable ``gluino $m_{T2}$ (stransverse mass)" which corresponds
to the Cambridge $m_{T2}$ variable \cite{mt201,mt202} applied to
pair produced gluinos each of which is decaying to two quarks and
one invisible neutralino LSP. Analytic expression of $m_{T2}^{\max}$
($=$ maximum of the gluino $m_{T2}$ over all events) as a function
of the {\it trial} LSP mass $m_\chi$ has been discussed with an
observation that $m_{T2}^{\rm max}$ has a {\it kink structure}, i.e.
a continuous but not differentiable cusp,  at $m_\chi=$ true LSP
mass, from which the gluino mass and the LSP mass can be determined
simultaneously with good accuracy. If squarks are lighter than
gluino, the gluino $m_{T2}^{\rm max}$ could determine the squark
mass also.  In this paper, we wish to provide a detailed discussion
of  $m_{T2}$ in more general context, including the features which
have been reported in \cite{mt203}\footnote{After \cite{mt203}, the
kink structure of the endpoint values of transverse mass observable
has been discussed also in \cite{gripaios,barr07}. For other
approaches to measure superparticle masses at hadron collider, see
\cite{lhclc,others1,others2,others3}.}.

This paper is organized as follows. In section 2, we discuss some
generic properties of $m_{T2}$,  and derive the analytic
expression of $m_{T2}^{\rm max}$ for general symmetric decay of
pair-produced mother superparticles. In section 3,  we consider
two specific processes, the decays of squark pair and of gluino
pair, to examine the structure  of $m_{T2}$ in somewhat detail,
and  also  perform a Monte Carlo LHC simulation for some
superparticle spectra to examine how well can $m_{T2}$ and
$m_{T2}^{\rm max}$ be constructed from real collider data. Section
4 is devoted to the conclusion.

\section{Generic features of $m_{T2}$}
\label{sec:generic_feature}

The kinematic variable `transverse mass' ($m_T$) has been introduced
to measure the $W$ boson mass from the decay $W \rightarrow l\nu$
\cite{mt}, for which  the transverse mass is given by
\begin{eqnarray}
m_T^2 = m_l^2 + m_\nu^2 + 2 (E^l_{T} E^\nu_{T} - {\bold p}^l_{T}
\cdot {\bold p}^\nu_{T}),
\end{eqnarray}
where $m_{l}, m_{\nu}$ and  ${\bold p}^l_{T}, \bold{p}^\nu_{T}$
denote the mass and transverse momentum of the corresponding
particle, respectively, and the transverse energies are defined as
\begin{eqnarray}
E^l_{T} = \sqrt{|{\bold p}^l_{T}|^2+ m_l^2},~~~ E^\nu_{T} =
\sqrt{|{\bold p}^\nu_{T}|^2+ m_\nu^2}.
\end{eqnarray}
On the other hand, the physical $W$ mass is given by
\begin{eqnarray} m_W^2
&=&m_l^2 + m_\nu^2 + 2p_l\cdot p_\nu \nonumber \\
&=& m_l^2 + m_\nu^2+2 (E^l_{T} E^\nu_{T}\cosh \Delta\eta - {\bold
p}^l_{T} \cdot {\bold p}^\nu_{T}) \,\geq\, m_T^2, \end{eqnarray}
where $\Delta \eta=\eta_l-\eta_\nu$ is the rapidity difference for a
4-momentum parameterized as $$p^\mu=(E_T\cosh \eta, \bold{p}_T,
E_T\sinh\eta).$$ Although the neutrino cannot be observed directly,
its transverse momentum can be inferred from the measured total
missing transverse momentum in event-by-event basis. Then, for each
event, the transverse mass of $W$ can be constructed from the
observed values of  $\bold{p}_T^l$ and $\bold{p}_T^\nu$, and an
endpoint measurement of the $m_T$ distribution determines the mother
particle mass $m_W$:
\begin{eqnarray}
m_W = \max_{\{\mbox{all events}\}} \big[\, m_T\,\big].
\end{eqnarray}

More challenging situation for  experimental measurement of unknown
mother particle mass would be the case when there are more than one
final state particles escaping the detection, so that the transverse
momentum of each invisible particle can {\it not} be determined
although the total missing transverse momentum is known. Additional
difficulty would arise if the mass of the invisible daughter
particle were not known in advance. Such situation is what one
actually encounters in supersymmetric extension of the SM  with
conserved $R$-parity, in which superparticles are pair produced in
collider experiment and each superparticle decay ends up with
producing an invisible lightest supersymmetric particle (LSP) with
unknown mass.

The $m_{T2}$ variable \cite{mt201,mt202} which is sometimes called
the ``stransverse mass" is a generalization of the transverse mass
to the case that a pair of massive mother particles are produced in
hadron collider with a vanishing total transverse momentum in the
laboratory frame, and subsequently  decay to  daughter particles
including ${two}$ invisible particles in the final state\footnote{In
Ref.\cite{mt202}, $m_{T2}$ has been further generalized to the case
involving more missing particles than two.}.
In this paper,  we will concentrate on the case that each mother
particle decays into the same set of daughter particles, since such
symmetric decay typically has higher event rate while showing the
non-trivial structure which will be discussed in the following. Fig.
1 shows  an example of such process in which mother superparticles
were pair-produced and each of them decays into one neutralino LSP
($\tilde{\chi}_1^0$) and some visible particles. While the invisible
part of each decay consists of only one particle (neutralino LSP),
the visible part might contain one or more visible particle(s) in
general.

\begin{figure}[ht!]
\begin{center}
\epsfig{figure=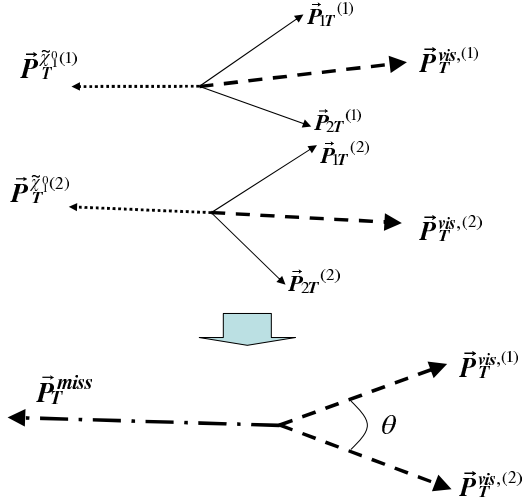,width=8cm,height=8cm}
\end{center}
\caption{Kinematic situation for $m_{T2}$ where ${\bf p}_T^{miss}$
denotes the total missing transverse momentum.}
\label{fig:kinematic}
\end{figure}


With two invisible LSPs in the final state, each LSP momentum can
not be determined although the total missing transverse momentum
$\bold{p}_T^{miss}$ can be measured experimentally. Furthermore, the
LSP mass might not be known in advance. In such situation, one  can
introduce  a {\it trial} LSP mass $m_\chi$, and define the $m_{T2}$
variable as follows \cite{mt201,mt202}:
\begin{equation}
m_{T2}(\mathbf{p}_{T}^{vis(1)}
,\,m_{vis}^{(1)},\,\mathbf{p}_{T}^{vis(2)},\,m_{vis}^{(2)},\,m_{\chi})\equiv
\min_{\{{\bf p}_T^{\chi(1)}+{\bf p}_T^{\chi(2)}=-{\bf
p}_T^{vis(1)}-{\bf p}_T^{vis(2)}\}} \left[
\mathrm{max}\{m_{T}^{(1)}, m_{T}^{(2)}\}\right], \label{mt2_def}
\end{equation}
where the minimization is performed over {\it trial} LSP momenta
${\bf p}_T^{\chi(i)}$ constrained as
$$
\bold{p}_T^{\chi(1)}+\bold{p}_T^{\chi(2)}=\bold{p}_T^{miss},$$ and
$m_T^{(i)}$ ($i=1,2$) denotes the transverse mass of the decay
product of each initial mother particle: \begin{align} m_{T}^{(i)}
& = \sqrt{(m_{vis}^{(i)})^2 +m_{\chi}^{2}+ 2(\,E_{T}^{vis(i)}
E_{T}^{\chi(i)}-\,\mathbf{p}_{T}^{vis(i)}\cdot\mathbf{p}_{T}^{\chi(i)})},
\end{align}
where $m_{vis}^{(i)}$ and $\,\mathbf{p}_{T}^{vis(i)}$ are the total
invariant mass and the total transverse momentum of the visible part
of each decay product:
\begin{eqnarray} \bold{p}^{vis{(i)}}_T&=&\sum_\alpha
\bold{p}_{{\alpha}T} ,\nonumber \\ (m_{vis}^{(i)})^2&=&\sum_\alpha
m_\alpha^2 + 2\sum_{\alpha>\beta}(E_{{\alpha}}E_{{\beta
}}-\bold{p}_{{\alpha}}\cdot \bold{p}_{{\beta}}), \end{eqnarray}
where $E_{\alpha}=\sqrt{m_\alpha^2+|\bold{p}_{\alpha}|^2}$,
$\bold{p}_\alpha$, and $\bold{p}_{\alpha T}$ denote the energy,
momentum, and transverse momentum, respectively,  of the $\alpha$-th
visible particle in the decay product of one mother particle, which
are measured in the laboratory frame. Here, we ignore the initial
state radiation effect, and then the total missing transverse
momentum is given by
\begin{eqnarray}
\bold{p}_T^{miss}=-(\bold{p}_T^{vis{(1)}}+\bold{p}_T^{vis{(2)}}),
\end{eqnarray}
and the transverse energies of the each visible system and of the
LSP are defined as
\begin{eqnarray}
E_{T}^{vis(i)}\equiv\sqrt{|\,\mathbf{p}_{T}^{vis(i)}|^{2}%
+(m_{vis}^{(i)})^2}, \quad E_{T}^{\chi(i)}\equiv
\sqrt{|{\mathbf{p}_{T}^{\chi(i)}}|^{2}+m_{\chi}^{2}}.
\end{eqnarray}
In fact, one can consider also an $m_{T2}$ with $m_{vis}^{(i)}$
replaced by the transverse mass of the visible part:
\begin{eqnarray}\big(m_T^{vis(i)}\big)^2=\sum_\alpha m_\alpha^2 +
2\sum_{\alpha>\beta}(E_{{\alpha}T}E_{{\beta
}T}-\bold{p}_{{\alpha}T}\cdot \bold{p}_{{\beta}T}),\end{eqnarray}
 where
$E_{\alpha T}=\sqrt{m_\alpha^2+|\bold{p}_{\alpha T}|^2}$. Since
${\bf p}_T^{vis(i)}$ and $m_{vis}^{(i)}$ are treated as independent
variables in the definition (\ref{mt2_def}), once one finds the
functional form of $m_{T2}$ in terms of ${\bf p}_T^{vis(i)}$ and
$m_{vis}^{(i)}$, the other $m_{T2}$ defined in terms of ${\bf
p}_T^{vis(i)}$ and $m_T^{vis(i)}$ can be easily obtained by
replacing $m_{vis}^{(i)}$ with $m_T^{vis(i)}$.


Even without knowing the transverse momentum of each LSP, one can
ensure that the true mother particle mass $\tilde{m}$ can  {\it not}
be smaller than $m_{T2}$ when the trial LSP mass $m_\chi$ is chosen
to be the true LSP mass $m_{\tilde\chi_1^0}$. This suggests that
$\tilde{m}$ might be able to be determined by the endpoint value of
$m_{T2}$ distribution:
\begin{eqnarray}
m_{T2}^{\rm max} (m_\chi) \equiv \max_{\{\mbox{all
events}\}}\left[\,m_{T2}(\mathbf{p}_{T}^{vis(1)}
,\,m_{vis}^{(1)},\,\mathbf{p}_{T}^{vis(2)},\,m_{vis}^{(2)},\,m_{\chi})
\,\right]
\end{eqnarray}
which is a function of the trial LSP mass $m_\chi$, satisfying
\begin{eqnarray} m_{T2}^{\rm
max}(m_\chi=m_{\tilde\chi_1^0})\,=\,\tilde{m}\,\equiv\, \mbox{mother
particle mass}.
\end{eqnarray}

In order to see how $m_{T2}$  is determined for a given event, let
us first consider the minimization of $m_T$
over unconstrained trial LSP momentum $\bold{p}_T^{\chi}$.
Differentiating $m_T^2$ by $\bold{p}_T^\chi$, one  finds
\begin{eqnarray}
{\partial m_T^2 \over \partial {\bold p}_T^\chi}=
2\left(E_T^{vis}{\bold p_T^\chi \over E_T^\chi}-{\bold
p_T^{vis}}\right).
\end{eqnarray}
This implies that $m_{T}$ has a stationary value when the trial LSP
momentum satisfies
\begin{eqnarray}
{\bold p}_T^\chi =\frac{E_T^\chi} {E_T^{vis}}{\bold
p}_T^{vis}=\frac{m_\chi}{m_{vis}}{\bold p}_T^{vis}.
\label{uncon_min}
\end{eqnarray}
This stationary point actually corresponds to the global minimum of
$m_T$ for given values  of $m_{vis}$ and $m_\chi$:
\begin{eqnarray}
\big(m_{T}\big)_{\rm min}=m_T\big|_{{\bold p}_T^\chi /m_\chi={\bold
p}_T^{vis} / m_{vis}}\,=\, m_{vis}+m_\chi,
\end{eqnarray}
which is called the {\it unconstrained minimum} of the transverse
mass \cite{mt202}.


\begin{figure}[ht!]
\begin{center}
\epsfig{figure=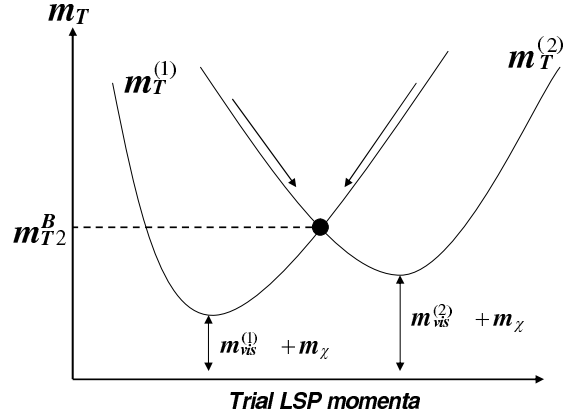,width=8cm,height=8cm}
\end{center}
\caption{A balanced $m_{T2}$ solution. } \label{fig:balanced}
\end{figure}

\begin{figure}[ht!]
\begin{center}
\epsfig{figure=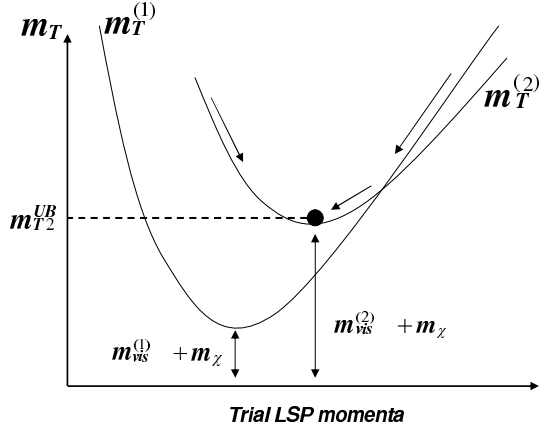,width=8cm,height=8cm}
\end{center}
\caption{An unbalanced $m_{T2}$ solution. } \label{fig:unbalanced}
\end{figure}


 With the above observation, one can consider two
different possibilities for how $m_T^{(i)}$ ($i=1,2$) depend on
trial LSP momenta, which are depicted in Fig. \ref{fig:balanced} and
Fig. \ref{fig:unbalanced}. The first possibility depicted in Fig.
\ref{fig:balanced} is that $m_T^{(i)}\geq m_T^{(j)}$  for both $i$
when the trial LSP transverse momenta take the value giving the
unconstrained minimum of $m_T^{(j)}$ ($j\neq i$), i.e.
\begin{eqnarray}
&&m_T^{(1)}\big|_{{\bf p}_T^{\chi(1)}=-{\bf p}_T^{vis(1)}-{\bf
p}_T^{vis(2)}-\tilde{\bf p}_T^{\chi(2)}}\,\geq\,
m_T^{(2)}\big|_{{\bf p}_T^{\chi(2)}=\tilde{\bf
p}_T^{\chi(2)}}\,=\,m_{vis}^{(2)}+m_\chi,\nonumber
\\
&&m_T^{(2)}\big|_{{\bf p}_T^{\chi(2)}=-{\bf p}_T^{vis(1)}-{\bf
p}_T^{vis(2)}-\tilde{\bf p}_T^{\chi(1)}}\,\geq\,
m_T^{(1)}\big|_{{\bf p}_T^{\chi(1)}=\tilde{\bf
p}_T^{\chi(1)}}\,=\,m_{vis}^{(1)}+m_\chi,\label{bal_region}
\end{eqnarray}
where $\tilde{\bf p}_T^{\chi(i)}$ is the trial LSP transverse
momentum  giving the unconstrained minimum of $m_T^{(i)}$:
$$\tilde{\bf p}_T^{\chi(i)}/m_\chi={\bf p}_T^{vis(i)}/m_{vis}^{(i)}.$$
 In such case,  the minimum of $\mbox{max}\{m_{T}^{(1)}, m_{T}^{(2)}\}$ over
possible trial LSP momenta is given by a {\it balanced} $m_{T2}$
solution \cite{mt202}
\begin{eqnarray}
m_{T2}^{\rm bal}= \min_{\{ {\bold p}_T^{\chi(1)}+{\bold
p}_T^{\chi(2)}=-{\bold p}_T^{{vis}(1)}-{\bold p}_T^{{vis}(2)},
\,m_T^{(1)}=m_T^{(2)}\}} \,\big[\,m_T^{(1)}\,\big],
\label{balanced_def}
\end{eqnarray}
where the minimization is performed over ${\bf p}_T^{\chi (i)}$
satisfying
\begin{eqnarray}
m_T^{(1)}({\bold p}_T^{{vis}(1)},{\bold
p}_T^{\chi(1)},m_{vis}^{(1)}, m_\chi)&=&m_T^{(2)}({\bold
p}_T^{{vis}(2)},{\bold p}_T^{\chi(2)},m_{vis}^{(2)},
m_\chi),\nonumber \\
{\bold p}_T^{\chi(1)}+{\bold p}_T^{\chi(2)}&=&-{\bold
p}_T^{{vis}(1)}-{\bold p}_T^{{vis}(2)}.
 \label{con_balanced}\end{eqnarray}

The condition (\ref{bal_region}) might not be satisfied as in the
case of Fig. \ref{fig:unbalanced} for which
\begin{eqnarray}
m_T^{(1)}\big|_{{\bf p}_T^{\chi(1)}=-{\bf p}_T^{vis(1)}-{\bf
p}_T^{vis(2)}-\tilde{\bf p}_T^{\chi(2)}}\,<\, m_T^{(2)}\big|_{{\bf
p}_T^{\chi(2)}=\tilde{\bf p}_T^{\chi(2)}}=m_{vis}^{(2)}+m_\chi,
\label{con_unbalanced}\end{eqnarray} where $\tilde{\bf
p}_T^{\chi(2)}$ is the trial LSP transverse momentum giving the
unconstrained minimum of $m_T^{(2)}$.
In such case, $m_{T2}$ is obviously given by the unconstrained
minimum of $m_T^{(2)}$. Solutions obtained in this way are called
the {\it unbalanced} $m_{T2}$ solution \cite{mt202}, and given by
\begin{eqnarray}
m^{\rm unbal}_{T2}= m_T^{(i)}\big|_{{\bold p}_T^{\chi(i)}
/m_\chi={\bold p}_T^{vis(i)} / m_{vis}^{(i)}}\,=\,
m_{vis}^{(i)}+m_\chi \quad (i=1 \,\, \mbox{or}\,\, 2).
\label{unbalanced}\end{eqnarray} Note that unbalanced  solution can
accompany a balanced solution as Fig. \ref{fig:unbalanced} for which
the crossing point of $m_T^{(1)}$ and $m_{T}^{(2)}$ corresponds to a
balanced $m_{T2}$ solution according to the definition
(\ref{balanced_def}). However obviously such balanced solution  can
{\it not} be a genuine $m_{T2}$ defined as (\ref{mt2_def}).

 Recently, analytic expression
of the balanced $m_{T2}$ solution for generic event has been derived
in Ref. \cite{mtgen}, which we briefly recapitulate in the
following.  For this, let us rewrite $m_{T2}$\footnote{Throughout
this paper, we ignore the initial state radiation, so the total
missing transverse momentum is given by
$\bold{p}_T^{miss}=-(\bold{p}_T^{vis{(1)}}+\bold{p}_T^{vis{(2)}})$.}
as
\begin{eqnarray}
m_{T2}^2 = \min_{\substack{\beta_1+\beta_2
=\sqrt{s}\Lambda-(\alpha_1+\alpha_2)
\\ \beta_1^2=\beta_2^2=m_\chi^2}}
\left[ \mathrm{max}\{(\alpha_1+\beta_1)^2
,(\alpha_2+\beta_2)^2\}\right],
\end{eqnarray}
where $V^2=(V^0)^2-\vec{V}\cdot\vec{V}$ is the scalar product of the
$(1+2)$-dimensional momentum vector $V^\mu=(V^0,\vec{V})$, and the
$(1+2)$-dimensional momenta $\alpha_i^\mu, \beta_i^\mu$ are given by
\begin{eqnarray}
&&\alpha_1^\mu=(E_T^{vis(1)},{\bold p}_T^{vis(1)}),\quad
\alpha_2^\mu=(E_T^{vis(2)}, {\bold p}_T^{vis(2)}),
\nonumber \\
&& \beta_1^\mu=(E_T^{{\chi}(1)},{\bold p}_T^{{\chi}(1)}),\quad
\beta_2^\mu=(E_T^{{\chi}(2)}, {\bold p}_T^{{\chi}(2)}).
\end{eqnarray}
These momenta are related as
\begin{equation}
\alpha_1^\mu + \beta_1^\mu + \alpha_2^\mu + \beta_2^\mu =\sqrt{s}\,
\Lambda^\mu= \sqrt{s}\,(1,0,0),
\end{equation}
where $\sqrt{s}$ corresponds to the total transverse  energy of the
event. The square of the balanced $m_{T2}$ solution  corresponds to
the minimum of $(\alpha_1+\beta_1)^2$ over all possible values of
$\sqrt{s}$, subject to the following constraint:
\begin{eqnarray}
\beta_1^2=\beta_2^2=m_\chi^2,\quad
(\alpha_1+\beta_1)^2=(\alpha_2+\beta_2)^2. \label{3con}
\end{eqnarray}
The procedure to obtain the balanced $m_{T2}$ solution is first to
solve (\ref{3con}) for $\beta_1^\mu$ and $\beta_2^\mu$, and next
minimize the resulting $(\alpha_1+\beta_1)^2$ over all allowed
values of $\sqrt{s}$. Then, one finds \cite{mtgen}
\begin{eqnarray}
 &&\big(m^{\rm bal}_{T2}\big)^2\,=\, m_\chi^2 + A_T
 \nonumber \\
 &&\qquad +\, \sqrt{\left(1+ {4
m_\chi^2 \over
2A_T-\left(m_{vis}^{(1)}\right)^2-\left(m_{vis}^{(2)}\right)^2}\right)
\left(A_T^2 -\Big(m_{vis}^{(1)}m_{vis}^{(2)}\Big)^2 \right)},
\label{balanced}
\end{eqnarray}
where $A_T$ is the Euclidean inner product of the two
(1+2)-dimensional visible momenta $\alpha_1^\mu$ and
$\alpha_2^\mu$:\begin{eqnarray}
A_T&\equiv&\alpha_1^0\alpha_2^0+\vec{\alpha_1}\cdot\vec{\alpha_2}
\nonumber
\\
&=&E_T^{{vis}(1)}E_T^{{vis}(2)}+{\bold p}_T^{{vis}(1)}\cdot {\bold
p}_T^{{vis}(2)},
\end{eqnarray}
and we have rewritten the result of \cite{mtgen} in a form
convenient for our later discussion.

The $m_{T2}$ solution obtained above has invariance properties which
will be useful for the derivation of the possible range of $m_{T2}$.
First of all,  the transverse masses $m_T^{(i)}$  and thus $m_{T2}$
are invariant under arbitrary {\it independent longitudinal Lorentz
boost} of each mother particle in the pair, which is obviously true
as both $E_T$ and ${\bf p}_T$ are invariant under longitudinal
boost.
 Furthermore, $m_{T2}$ is invariant also under the following
transformation of the $(1+2)$-dimensional momenta
$\alpha_i^\mu=(E_T^{vis (i)}, {\bf p}_T^{vis(i)})$ and
$\beta_i^\mu=(E_T^{\chi(i)},{\bf p}_T^{\chi(i)})$:
\begin{eqnarray}
&&\alpha_1^\mu\rightarrow \Lambda^\mu_\nu(\vec{v})\alpha_1^\nu,\quad
\beta_1^\mu\rightarrow \Lambda^\mu_\nu(\vec{v})\beta_1^\nu,\nonumber \\
&&\alpha_2^\mu\rightarrow \Lambda^\mu_\nu(-\vec{v})\alpha_2^\nu,
\quad \beta_2^\mu\rightarrow \Lambda^\mu_\nu(-\vec{v})\beta_2^\nu,
\label{tr_boost}
\end{eqnarray}
where $\Lambda^\mu_\nu(\vec{v})$ denotes the $(1+2)$-dimensional
Lorentz transformation matrix for 2-dimensional boost parameter
$\vec{v}$. The condition (\ref{bal_region}) and the relation
(\ref{con_balanced}) are covariant under the above transformation,
i.e. if (\ref{bal_region}) or (\ref{con_balanced}) is satisfied by
some $(1+2)$-dimensional momenta, it is satisfied also by the
transformed momenta. (In appendix A, we provide a more detailed
discussion on the covariance of (\ref{bal_region}) and
(\ref{con_unbalanced}) under the transformation (\ref{tr_boost}).)
As the transverse mass itself is obviously invariant, this means
that the balanced solution $m^{\rm bal}_{T2}$ defined as
(\ref{balanced_def}) is invariant under the transformation
(\ref{tr_boost}), which can be confirmed also by the explicit
solution (\ref{balanced}). As for the unbalanced solution,
(\ref{uncon_min}) and (\ref{con_unbalanced}) are covariant, so the
unbalanced solution  $m_{T2}^{\rm unbal}$ defined as
(\ref{unbalanced}) is invariant also.

In fact, for generic  ${\bf p}^{vis(i)}$, the  transformation
(\ref{tr_boost}) of the $(1+2)$-dimensional vector $(E_T,{\bf p}_T)$
does {\it not} have a direct connection   to a true Lorentz
transformation of the 4-momentum $(E,{\bf p})$. However, if both
${\bf p}^{vis(1)}$ and ${\bf p}^{vis(2)}$ are in transverse
direction, it can be identified as a true Lorentz transformation
corresponding to {\it a back-to-back boost of the mother particle
pair in transverse direction}.  We thus conclude that $m_{T2}$ for
visible momenta satisfying ${\bf p}^{vis(i)}={\bf p}_T^{vis(i)}$
($i=1,2$) is invariant under a back-to-back boost of the mother
particle pair in the direction along  the transverse plane $T$.

Let us now examine the possible range of $m_{T2}$, in particular its
maximal value. Our starting point is the following theorem which
will be proved in  appendix A: $m_{T2}$ of any event induced by
mother particle pair having a vanishing total transverse momentum in
the laboratory frame is {\it bounded above} by another $m_{T2}$ of
an event induced by  mother particle pair {\it at rest}. More
explicitly, for generic ${\bf p}^{vis (i)}$ measured in the
laboratory frame,
\begin{eqnarray}
m_{T2}({\bf p}_T^{vis(i)},m_{vis}^{(i)},m_\chi) \,\leq \,
m_{T^{\prime}2}({\bf q}^{vis(i)},m_{vis}^{(i)},m_\chi),
\label{theorem}
\end{eqnarray}
where  ${\bf q}^{vis(i)}$ is the Lorentz boost of ${\bf p}^{vis(i)}$
to the rest frame of the $i$-th mother particle, $T^{\prime}$ is the
plane spanned by ${\bf q}^{vis(1)}$ and ${\bf q}^{vis(2)}$, so ${\bf
q}^{vis(i)}={\bf q}_{T^\prime}^{vis(i)}$ by definition, and the
equality in the above bound holds when $T=T^\prime$. Note that $T$
is a fixed transverse plane which is independent of the event
momenta, while $T^\prime$ varies following the rest frame momenta
${\bf q}^{vis(i)}$.

By definition, ${\bf q}^{vis(i)}$ corresponds to the total visible
momentum for the decay  of the $i$-th mother particle,
$\Phi_{i}\rightarrow \mbox{visibles}+ \tilde{\chi}_1^0$, measured in
the rest frame of  $\Phi_{i}$, and thus its magnitude is uniquely
fixed by the mother particle mass $\tilde{m}$, the invariant mass
$m_{vis}^{(i)}$ of the visible part, and the true LSP mass
$m_{\tilde{\chi}_1^0}$:
\begin{eqnarray}
|{\bf q}^{vis(i)}|&=&\frac{1}{2
\tilde{m}}\left[\Big((\tilde{m}+m_{vis}^{(i)})^2-m_{\tilde\chi_1^0}^2\Big)
\Big((\tilde{m}-m_{vis}^{(i)})^2-m_{\tilde\chi_1^0}^2\Big)\right]^{1/2}.
 \label{relation}
\end{eqnarray}
As a consequence, $m_{T^{\prime}2}$ can be described by the three
event variables,  $m_{vis}^{(1)}, m_{vis}^{(2)}$ and $\theta$, where
$\theta$ is the angle between ${\bf q}^{vis(1)}$ and ${\bf
q}^{vis(2)}$:
\begin{eqnarray}
m_{T^{\prime}2}({\bf
q}^{vis(i)},m_{vis}^{(i)},m_\chi)\,\equiv\,{\cal F}(m_{vis}^{(i)},
\theta, m_\chi).
\end{eqnarray}
 Combined with (\ref{theorem}), this leads to
\begin{eqnarray}
m_{T2}^{\rm max}(m_\chi) &\equiv& \max_{\{\mbox{all events in the
lab frame}\}} \big[m_{T2}({\bf
p}_T^{vis(i)},m_{vis}^{(i)},m_\chi)\big]
\nonumber \\
 &\leq & \max_{\{m_{vis}^{(i)},\theta\}} \big[ {\cal
F}(m_{vis}^{(i)},\theta, m_\chi)\big] \equiv {\cal F}^{\rm
max}(m_\chi). \label{bound_max}
\end{eqnarray}

It is always a possible event that both ${\bf q}^{vis(1)}$ and ${\bf
q}^{vis(2)}$ are along the direction of the transverse plane $T$,
i.e. $T^\prime=T$. For such event, the invariance of $m_{T2}$ under
longitudinal and back-to-back transverse boosts implies
\begin{eqnarray}
m_{T2}({\bf q}_T^{vis(i)}, m_{vis}^{(i)},m_\chi)= m_{T2}({\bf
p}_T^{vis(i)}, m_{vis}^{(i)},m_\chi)
\end{eqnarray}
where now ${\bf p}^{vis(i)}$  are the momenta obtained by arbitrary
back-to-back transverse boost along $T=T^\prime$ and subsequent
arbitrary longitudinal boost  starting from ${\bf q}^{vis(i)}$. For
any ${\bf q}^{vis(i)}$, one can choose an appropriate boost for
which ${\bf p}^{vis(i)}$ correspond to the visible momenta of some
event observed in the laboratory frame.
 This means that for any ${\bf q}^{vis(i)}$, there exist some laboratory events whose $m_{T2}$
 is same as the corresponding ${\cal F}(m_{vis}^{(i)},
\theta,m_\chi)$, therefore the bound (\ref{bound_max}) actually
corresponds to an identity as
\begin{eqnarray}
m_{T2}^{\rm max}(m_\chi) \,=\, {\cal F}^{\rm max}(m_\chi).
\label{mt2_max}
\end{eqnarray}

 In appendix B, we will show
\begin{eqnarray}
&&\frac{\partial {\cal F}}{\partial\theta}\leq 0 \quad\,\,\mbox{for
any}\,\, m_{vis}^{(i)},m_\chi,\,\,\mbox{and}\,\,0\leq\theta\leq\pi, \nonumber \\
&&\left.\frac{\partial {\cal F}}{\partial
m_{vis}^{(i)}}\right|_{\theta=0}=\left\{\begin{array}{ll}\,\leq
0\quad \mbox{for} \,\, m_{\chi}<m_{\tilde{\chi}_1^0}\,\,\mbox{and
any}\,\, m_{vis}^{(i)}\\ \,\geq 0 \quad \mbox{for} \,\,
m_{\chi}>m_{\tilde{\chi}_1^0} \,\,\mbox{and any}\,\, m_{vis}^{(i)},
\end{array}\right.
\label{gra}\end{eqnarray} and thus the global maximum of ${\cal F}$
over the full 3-dimensional event space of
$\{m_{vis}^{(i)},\theta\}$ is given by
\begin{eqnarray}
{\cal F}^{\rm max}(m_{\chi})=\left\{\begin{array}{ll} {\cal F}^{\rm
max}_<
  \quad \mbox{for} \,\,
m_{\chi}<m_{\tilde{\chi}_1^0} \\ {\cal F}^{\rm max}_{>}\quad
\mbox{for} \,\, m_{\chi}>m_{\tilde{\chi}_1^0},
\end{array}\right. \end{eqnarray}
where \begin{eqnarray}
 {\cal F}^{\rm max}_< &=&{\cal
F}(m_{vis}^{(1)}=m_{vis}^{\rm min}, m_{vis}^{(2)}=m_{vis}^{\rm
min},\theta=0,m_\chi),\nonumber \\{\cal F}^{\rm max}_{>} &=&{\cal
F}(m_{vis}^{(1)}=m_{vis}^{\rm max}, m_{vis}^{(2)}=m_{vis}^{\rm
max},\theta=0,m_\chi)
\end{eqnarray}
with \begin{eqnarray} m_{vis}^{\rm min}\leq m_{vis}^{(i)}\leq
m_{vis}^{\rm max}.\end{eqnarray} In this section, for simplicity we
limit ourselves to the one-dimensional event space with
$m_{vis}^{(1)}=m_{vis}^{(2)}$ and $\theta=0$, while leaving the
discussion of full event space in appendix B.

 For $m_{vis}^{(1)}=m_{vis}^{(2)}$, the unconstrained minima of
$m_T^{(i)}$ have the same value, so  $m_{T2}$ is always obtained as
a balanced solution. One can then  use the balanced solution
(\ref{balanced}) for ${\bf p}^{vis(i)}={\bf q}^{vis(i)}$  to find
\begin{eqnarray}\tilde{\cal F}(m_{vis},m_\chi)
&\equiv&{\cal
F}(m_{vis}^{(1)}=m_{vis},m_{vis}^{(2)}=m_{vis},\theta=0,m_\chi)
\nonumber \\
&=& \frac{\tilde{m}^2+m_{vis}^2-m_{\tilde{\chi}^0_1}^2}{2\tilde{m}}
+
\frac{\left[\big(\tilde{m}^2-m_{vis}^2+m_{\tilde{\chi}^0_1}^2\big)^2+
4\tilde{m}^2\big(m_\chi^2-m_{\tilde{\chi}^0_1}^2\big)\right]^{1/2}}{2\tilde{m}},
\end{eqnarray}
where $\tilde{m}$ is the mother particle mass,
$m_{\tilde{\chi}_1^0}$ is the true LSP mass, and we have used
(\ref{relation}) for $|{\bf q}^{vis(i)}|$.
 Note
that $\tilde{\cal
F}(m_{vis},m_\chi=m_{\tilde{\chi}_1^0})=\tilde{m}$, and thus
$m_{T2}^{\rm max}(m_\chi=m_{\tilde{\chi}_1^0})=\tilde{m}$ as
required.

The function $\tilde{\cal F}(m_{vis},m_\chi)$ has an interesting
feature. From
\begin{eqnarray}
{\partial \tilde{\cal F} \over \partial m_{vis}} ={m_{vis} \over
\tilde{m}} \left(1-{ \tilde{m}^2+m_{\tilde\chi_1^0}^2-(m_{vis})^2
\over \sqrt{(\tilde{m}^2+m_{\tilde\chi_1^0}^2-(m_{vis})^2)^2 +4
\tilde{m}^2 (m_\chi^2-m_{\tilde\chi_1^0}^2)}} \right),
\end{eqnarray}
one easily finds
\begin{eqnarray}
{\partial \tilde{\cal F} \over \partial m_{vis}}= \left\{
\begin{array}{ll}
\, \leq \, 0 \quad {\rm
if} ~~m_\chi < m_{\tilde\chi_1^0}\\
\,\geq\, 0 \quad {\rm if} ~~m_\chi
> m_{\tilde\chi_1^0}.
\end{array}\right.
\end{eqnarray}
This corresponds to the special limit of the general result
(\ref{gra}), and yields
\begin{eqnarray} m_{T2}^{\rm max}(m_\chi)
= \left \{\begin{array}{ll}\, {\cal F}^{\rm
max}_<(m_\chi)=\tilde{\cal F}(m_{vis}=m_{vis}^{\rm min},m_\chi) &
\,\,\mbox{if \,$m_\chi <
m_{\tilde\chi_1^0}$},\\
\, {\cal F}^{\rm max}_>(m_\chi)=\tilde{\cal F}(m_{vis}=m_{vis}^{\rm
max},m_\chi)& \,\,\mbox{if\, $m_\chi
> m_{\tilde\chi_1^0}$}.
\end{array}
\right. \label{mt2max}
\end{eqnarray}


In the above, we have obtained $m_{T2}^{\rm max}$ from the balanced
solution for $m_{vis}^{(1)}=m_{vis}^{(2)}$.
 However, this does not mean that $m_{T2}^{\rm max}$ is obtained
only by the balanced solution. As we will see in the next section,
$m_{T2}^{\rm max}$ for $m_\chi>m_{\tilde\chi_1^0}$ can be obtained
also by unbalanced $m_{T2}$ solution. In some of such case, the
balanced solution giving $m_{T2}^{\rm max}$ has ${\bf
p}_T^{miss}=0$, and thus is eliminated by the event selection
imposing a nonzero lower bound on $|{\bf p}_T^{miss}|$ when one
constructs $m_{T2}^{\rm max}$ from collider data. On the other hand,
the unbalanced solution giving $m_{T2}^{\rm max}$  has a large ${\bf
p}_T^{miss}$, so plays an crucial role for the construction of
$m_{T2}^{\rm max}$ from collider data.

If the decay product of each mother particle contains only one
visible particle $\psi$, then $m_{vis}$ is fixed to be $m_\psi$, and
 $m_{T2}^{\rm max}$ is given by
\begin{eqnarray} && m_{T2}^{\rm max}(m_\chi)=\tilde{\cal F}(m_{vis}=m_\psi,
m_\chi) \nonumber \\
&=&\frac{\tilde{m}^2+m_{\psi}^2-m_{\tilde{\chi}^0_1}^2}{2\tilde{m}}
+
\frac{\left[\big(\tilde{m}^2-m_\psi^2+m_{\tilde{\chi}^0_1}^2\big)^2+
4\tilde{m}^2\big(m_\chi^2-m_{\tilde{\chi}^0_1}^2\big)\right]^{1/2}}{2\tilde{m}}.
\label{onevis}
\end{eqnarray}
 Usually $\psi$ is much lighter than the LSP, so
one can take the approximation $m_\psi\simeq 0$, for which
\begin{eqnarray} m_{T2}^{\rm max}(m_\chi) &=&\frac{\tilde{m}^2-m_{\tilde{\chi}^0_1}^2}{2\tilde{m}}+
\sqrt{\left(\frac{\tilde{m}^2-m_{\tilde{\chi}^0_1}^2}{2\tilde{m}}\right)^2+m_\chi^2}.
\label{mt2_single}
\end{eqnarray}
 A simple
example  giving this form of $m_{T2}^{\rm max}$  would be the squark
pair decay\footnote{Throughout this paper, we do not distinguish
squark from anti-squark as we are considering only the kinematics.},
$\tilde{q}\tilde{q}\rightarrow q\tilde{\chi}^0_1
{q}\tilde{\chi}^0_1$, whose $m_{T2}$ will be discussed in more
detail in the next section.

If the decay product of each mother particle contains more than one
visible particles,  $m_{T2}^{\rm max}$ has an interesting feature
which was first noticed in \cite{mt203}. In such case, $m_{vis}$ can
vary from $m_{vis}^{\rm min}$ to $m_{vis}^{\rm max}$, and so
 $m_{T2}^{\rm max}(m_\chi)$ at $m_\chi> m_{\tilde{\chi}_1^0}$ takes a {\it different} functional form
from the one at $m_\chi < m_{\tilde{\chi}_1^0}$.
 As a consequence,
$m_{T2}^{\max}$ has a {\it kink structure}, i.e. a continuous but
not differentiable cusp, at $m_\chi=m_{\tilde{\chi}^0_1}$:
\begin{eqnarray}
\frac{({d{\cal F}_>^{\rm
max}}/{dm_\chi})_{m_\chi=m_{\tilde{\chi}^0_1}}}{({d{\cal F}_<^{\rm
max}}/{dm_\chi})_{m_\chi=m_{\tilde{\chi}^0_1}}}
\,=\,1+\frac{(m_{vis}^{\rm max})^2-(m_{vis}^{\rm
min})^2}{\tilde{m}^2+m_{\tilde{\chi}_1^0}^2-(m_{vis}^{\rm max})^2}
\,>\, 1, \label{cusp1}
\end{eqnarray}
which becomes sharper when $m_{vis}^{\rm max}$ becomes larger for a
given value of $m_{vis}^{\rm min}$.

If visible particles in the decay product are much lighter than LSP,
one can set $m_{vis}^{\rm min}\simeq 0$, and then
\begin{eqnarray} &&  {\cal F}^{\rm
max}_<(m_\chi)=\tilde{\cal F}(m_{vis}=0, m_\chi) \nonumber
\\
&=&\frac{\tilde{m}^2-m_{\tilde{\chi}^0_1}^2}{2\tilde{m}}+
\sqrt{\left(\frac{\tilde{m}^2-m_{\tilde{\chi}^0_1}^2}{2\tilde{m}}\right)^2+m_\chi^2}.
\label{multivis1}
\end{eqnarray}
 On the other hand, even for a given number of visible particles in the final state, $m_{T2}^{\rm max}$ can have
 a  different value depending upon the intermediate stage of the
decay process. If the decay process $\Phi_i\rightarrow
\mbox{visibles}+ \tilde{\chi}_1^0$  does not involve any
intermediate on-shell particle lighter than the mother particle
$\Phi_i$, one finds
\begin{eqnarray}m_{vis}^{\rm max}=
\tilde{m}-m_{\tilde{\chi}_1^0},\end{eqnarray} which results in
\begin{eqnarray}
 {\cal F}^{\rm
max}_>(m_\chi) \,=\,m_\chi +\left(\tilde{m}
-m_{\tilde\chi_1^0}\right). \label{multivis2}
\end{eqnarray}
However, if visible particles are produced by a chain of decay
processes involving intermediate on-shell particle(s), $m_{vis}^{\rm
max}$ has a smaller value.
 For instance,
if the decay chain involves one intermediate on-shell particle
$\phi$ with $m_\phi < \tilde{m}$, e.g. $\Phi_i\rightarrow
\phi+\mbox{visible}\rightarrow \tilde{\chi}_1^0+\mbox{more
visibles}$, the maximal value of $m_{vis}^{(i)}$ for the final state
 is given by
\begin{eqnarray} m_{vis}^{\rm max}={\sqrt{(\tilde{m}^2 - m_\phi^2)
(m_\phi^2 - m_{\tilde\chi_1^0}^2)} \over m_\phi},\end{eqnarray} for
which
\begin{eqnarray} {\cal F}^{\rm max}_>(m_\chi) &=& \left({\tilde{m}
\over 2} (1-{m_{\phi}^2 \over \tilde{m}^2}) +{\tilde{m} \over 2}
(1-{m_{\tilde\chi_1^0}^2 \over m_{\phi}^2})\right)\nonumber \\
&+&\sqrt{\left({\tilde{m} \over 2} (1-{{m_\phi}^2 \over
\tilde{m}^2}) -{\tilde{m} \over 2} (1-{m_{\tilde\chi_1^0}^2 \over
m_\phi^2})\right)^2+m_\chi^2}. \label{multivis3}
\end{eqnarray}

\begin{figure}[ht!]
\begin{center}
\epsfig{file=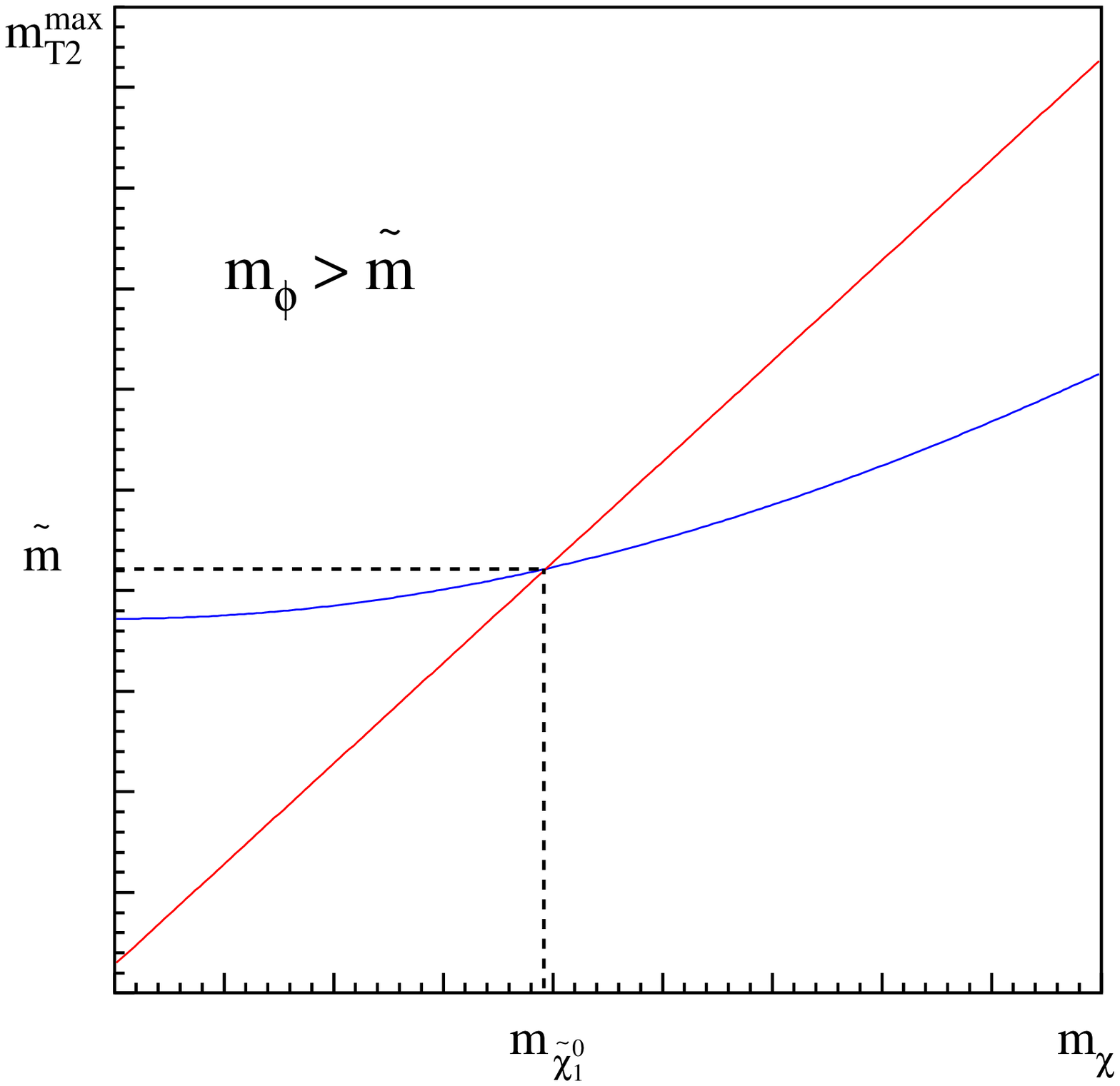, width=6cm, height=6cm}
\epsfig{file=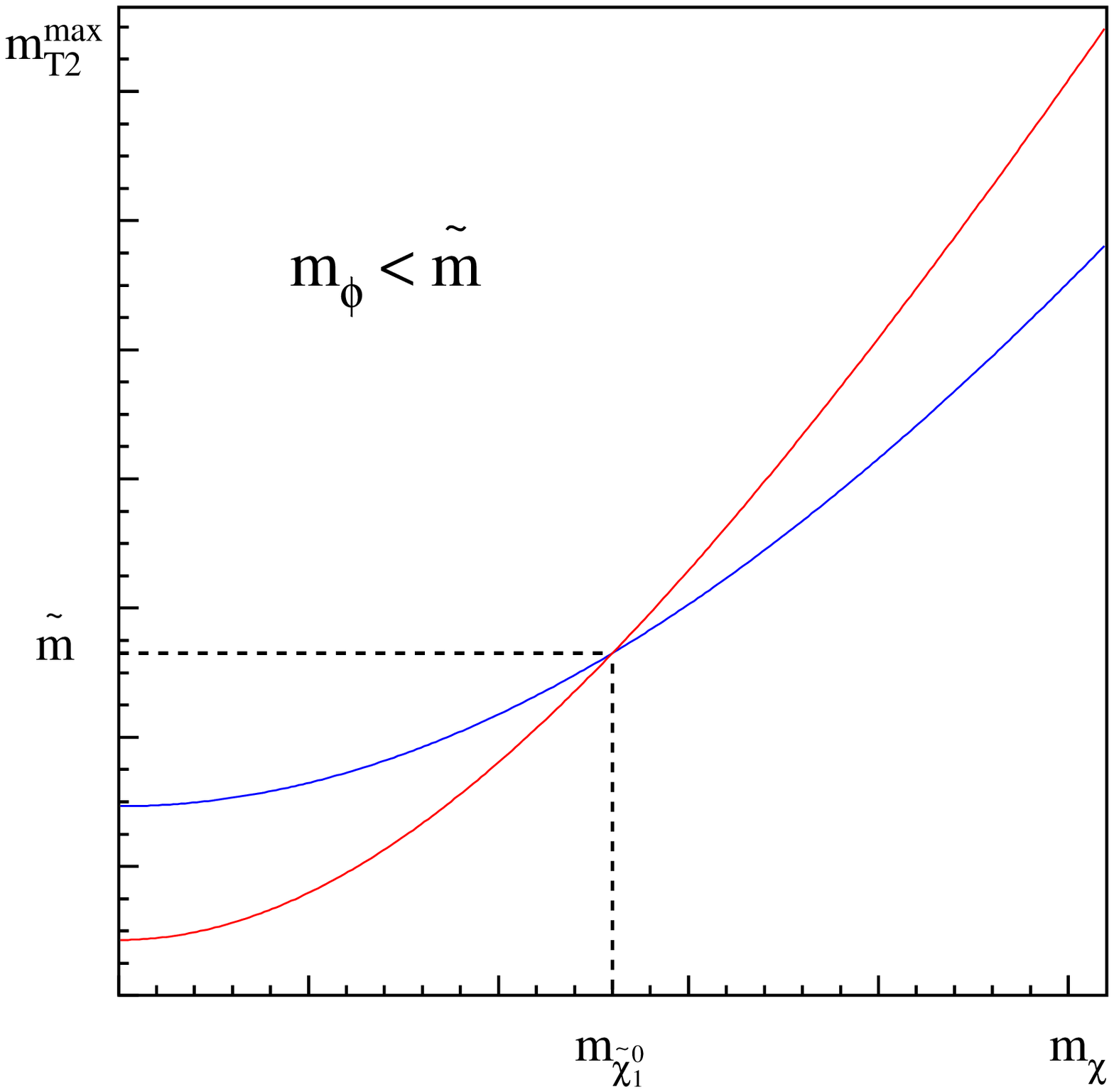, width=6cm, height=6cm}
\end{center}
\caption{$m_{T2}^{\rm max}$ for (a) $m_\phi > \tilde{m}$, (b)
$m_\phi < \tilde{m}$} \label{fig:mt2_heavy_lightsquark}
\end{figure}

In Fig. \ref{fig:mt2_heavy_lightsquark}, we depict the behavior of
$m_{T2}^{\rm max}(m_\chi)$ for the case without any intermediate
on-shell particle ($m_\phi >\tilde{m}$) and also the case of decay
chain involving an intermediate on-shell particle ($m_\phi <
\tilde{m}$). As the value of $m_{vis}^{\rm max}$ for the case
without intermediate on-shell particle is bigger than the value of
$m_{vis}^{\rm max}$ for the other case, the kink structure of Fig.
\ref{fig:mt2_heavy_lightsquark}(a) is sharper than that of Fig.
\ref{fig:mt2_heavy_lightsquark}(b) as anticipated in (\ref{cusp1}).
A simple example of the process that each mother particle is
producing more than one visible particles is the gluino pair decay:
$\tilde{g}\tilde{g}\rightarrow qq\tilde{\chi}^0_1
qq\tilde{\chi}^0_1$, whose $m_{T2}$ will be discussed  in more
detail in  the next section.

\section{Features of squark and gluino $m_{T2}$}
\label{sec:squark_mt2}

In this section, we discuss in more detail the $m_{T2}$ of two
specific processes, the decay of pair-produced squarks, ${\tilde
q}{\tilde q} \rightarrow q {\tilde\chi_1^0} q {\tilde\chi_1^0}$, as
an example of the case that each mother particle decays to one
visible particle and one invisible LSP, and the decay of
pair-produced gluinos, ${\tilde g}{\tilde g} \rightarrow qq
{\tilde\chi_1^0} qq {\tilde\chi_1^0}$, as an example of the next
case that the decay product of each mother particle contains more
than one visible particle. We also perform a Monte Carlo LHC
simulation for some superparticle spectra to examine how well can
$m_{T2}$ and $m_{T2}^{\rm max}$ be constructed from real collider
data.


\subsection{Squark $m_{T2}$}

Let us consider $m_{T2}$ for the process in which a pair of squarks
are produced in proton-proton collision and each squark decays
subsequently into one quark and one LSP:
\begin{eqnarray}
pp \rightarrow {\tilde q}{\tilde q} \rightarrow q {\tilde\chi_1^0} q
{\tilde\chi_1^0},
\end{eqnarray}
where $q$ denotes the 1st or 2nd generation quark. A characteristic
feature of this process
is that $m_{vis}^{(1)}=m_{vis}^{(2)}=m_q$, and thus $m_T^{(1)}$ and
$m_T^{(2)}$ have the same unconstrained minimum:
\begin{eqnarray} \big(m_{T}^{(1)}\big)_{\rm
min}=\big(m_{T}^{(2)}\big)_{\rm min}=m_{\chi}+m_q. \end{eqnarray}
Then, $m_{T2}$ is always obtained as a balanced solution as shown in
Fig. \ref{fig:balanced}, and the resulting balanced solution
(\ref{balanced})  is simplified as
\begin{eqnarray}
m_{T2}^2 &=& m_\chi^2 + A_T
+ \sqrt{\Big(A_T-m_q^2+2 m_\chi^2\Big) \Big(A_T+m_q^2\Big)},
\label{squark}
\end{eqnarray}
giving
\begin{eqnarray}
m_{T2} = \sqrt{{A_T+m_q^2 \over 2}}+ \sqrt{{A_T-m_q^2+2 m_\chi^2
\over 2}}, \label{sqmt2}
\end{eqnarray}
where
\begin{eqnarray}
A_T &\equiv& E_T^{{vis}(1)}E_T^{{vis}(2)}+{\bold
p}_T^{{vis}(1)}\cdot {\bold p}_T^{{vis}(1)}\nonumber \\
&=& E_T^{vis(1)} E_T^{vis(2)}+|{\bold p}_T^{vis(1)}||{\bold
p}_T^{vis(2)}|\, \cos\theta \label{euclidean}
\end{eqnarray}
for $E_T^{vis(i)} = \sqrt{|{\bold p}_T^{vis(i)}|^2 + m_q^2}$
$~(i=1,2)$.

As was discussed in the previous section, $m_{T2}$ of any event
induced by mother particle pair having a vanishing total transverse
momentum in the laboratory frame is bounded above by another
$m_{T2}$ of an event induced by mother particle pair at rest. For a
squark pair at rest, the magnitude of quark momentum from each
squark decay is given by
\begin{eqnarray}
|{\bold p}^{vis(i)}|= \frac{1}{2 m_{\tilde q}}\left[\big((m_{\tilde
q}+m_q)^2 - m_{\tilde\chi_1^0}^2 \big)\big((m_{\tilde q}-m_q)^2 -
m_{\tilde\chi_1^0}^2 \big)\right]^{1/2}.
\end{eqnarray}
It is then straightforward to find that $m_{T2}$ has a maximum  at
$\theta=0$ for visible momenta in transverse direction,
giving
\begin{eqnarray}
m_{T2}^{\rm max} &=&{m_{\tilde q}^2-m_{\tilde\chi_1^0}^2+m_q^2 \over
2 m_{\tilde q}} \nonumber \\
&+& \sqrt{{\big((m_{\tilde q}+m_q)^2 - m_{\tilde\chi_1^0}^2
\big)\big((m_{\tilde q}-m_q)^2 - m_{\tilde\chi_1^0}^2 \big) \over
4m^2_{\tilde q}}+m_\chi^2}
 \label{sqmt2max}
\end{eqnarray}
as obtained in (\ref{onevis}).
 In the limit when the quark mass is negligible, this expression
of $m^{\rm max}_{T2}$ is further simplified as
\begin{eqnarray}
m_{T2}^{\rm max} (m_\chi)= {m_{\tilde q}^2-m_{\tilde\chi_1^0}^2
\over 2 m_{\tilde q}} +\sqrt{\left({m_{\tilde
q}^2-m_{\tilde\chi_1^0}^2 \over 2 m_{\tilde q}}\right)^2 +
m_\chi^2}. \label{sqmt2max2}
\end{eqnarray}


As the above $m_{T2}^{\rm max}$ has been obtained from a momentum
configuration in which the two quarks produced by a squark pair at
rest are moving in the same direction, one might worry that
such two quarks might not be identified as two separate jets in real
collider data with realistic jet reconstruction. However, the quark
momenta from the  back-to-back  transverse boosted squark pair are
generically not in the same direction, and thus the boosted events
can provide two separate jets, while giving the same value of
$m_{T2}^{\rm max}$.

\begin{figure}[ht!]

\begin{center}
\epsfig{figure=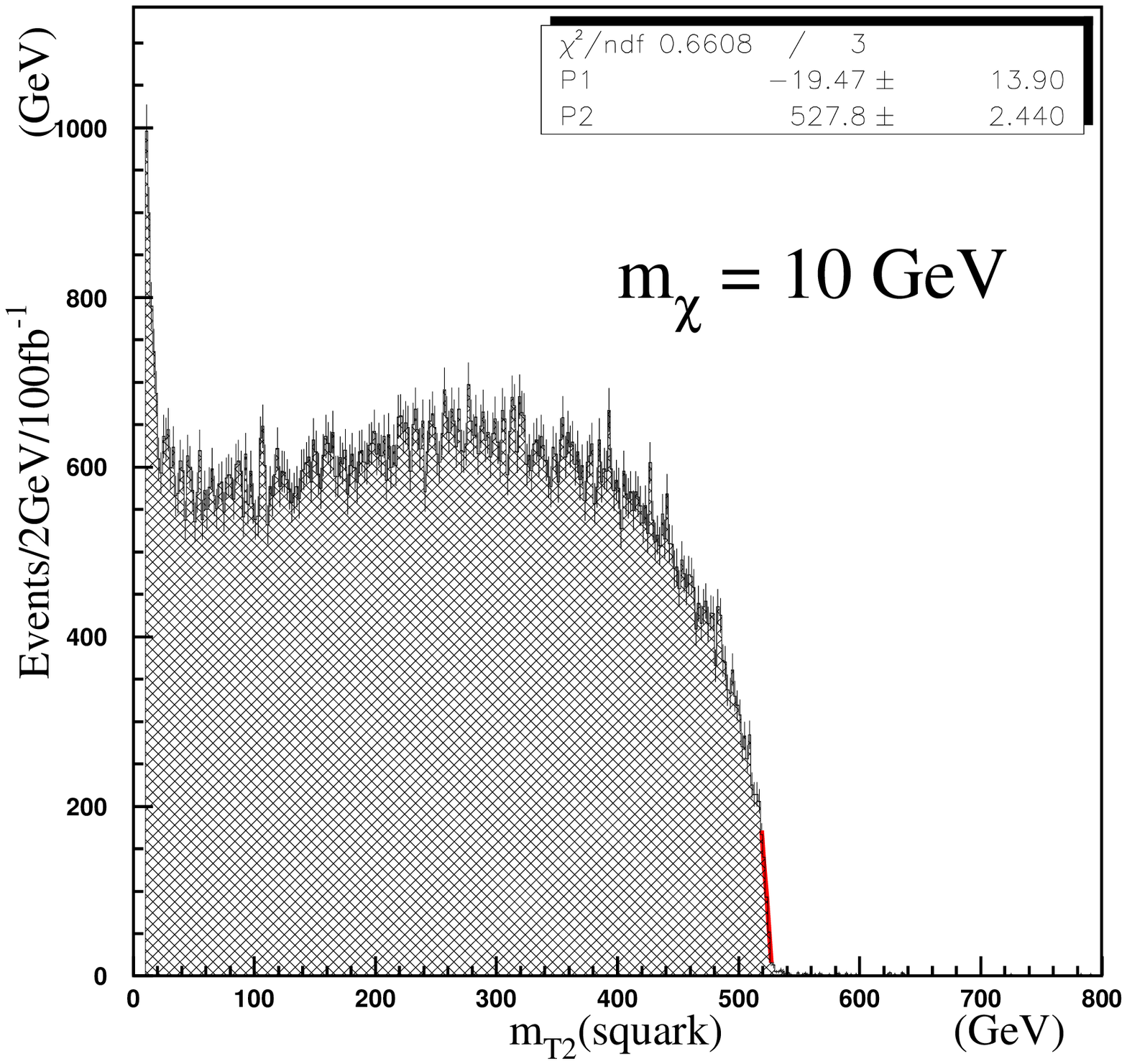,width=7cm,height=7cm}
\epsfig{figure=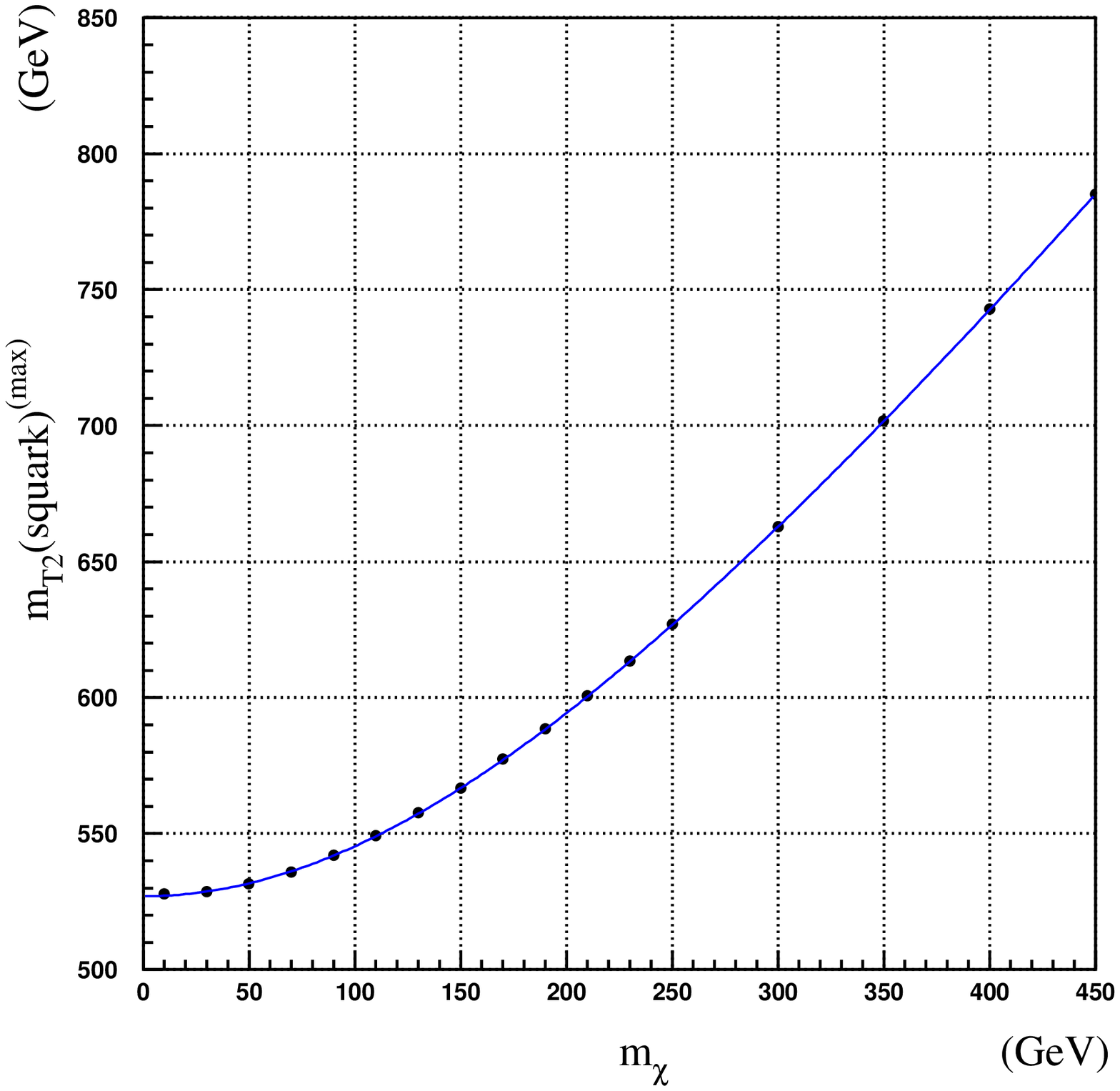,width=7cm,height=7cm}
\end{center}

\caption{(a) $m_{T2}$ distribution with $m_\chi=10$ GeV for the
mirage mediation parameter point (\ref{sqmt2-mass}). (b) Resulting
$m_{T2}^{\rm max}$ as a function of the trial LSP mass $m_\chi$.}
\label{fig:mirage}

\end{figure}

To see explicitly some features of the squark $m_{T2}$, we have
performed a Monte Carlo analysis for a SUSY parameter point in
mirage mediation scenario \cite{mirage}, which gives
\begin{eqnarray}
m_{\tilde q} = 697~{\rm GeV},~~m_{\tilde\chi_1^0}=344~{\rm GeV}.
\label{sqmt2-mass}
\end{eqnarray}
A Monte Carlo event sample for the signal $pp \rightarrow {\tilde q}
{\tilde q} \rightarrow q \tilde\chi_1^0 q \tilde\chi_1^0$ has been
generated in partonic-level using the PYTHIA event generator
\cite{pythia}. The $m_{T2}$ values for the event sample were then
calculated with a numerical code \cite{lester-code} implementing the
minimization over the trial LSP momenta. Fig. \ref{fig:mirage}(a)
shows the resulting $m_{T2}$ distribution for a trial LSP mass
$m_\chi=10$ GeV.
The distribution shows a sharp edge at $m_{T2}=527.8$ GeV, which is
very close to $m_{T2}^{\rm max}=527.4$ GeV obtained from the
analytic formula (\ref{sqmt2max2}).
Finally, $m_{T2}^{\rm max}$ as a function of the trial LSP mass
$m_\chi$ is shown in Fig. \ref{fig:mirage}(b). Here, the blue curve
represents the analytic formula  (\ref{sqmt2max2}), while the black
dots are obtained from the Monte Carlo data, which fit very well the
analytic curve.

\begin{figure}[ht!]
\begin{center}
\epsfig{figure=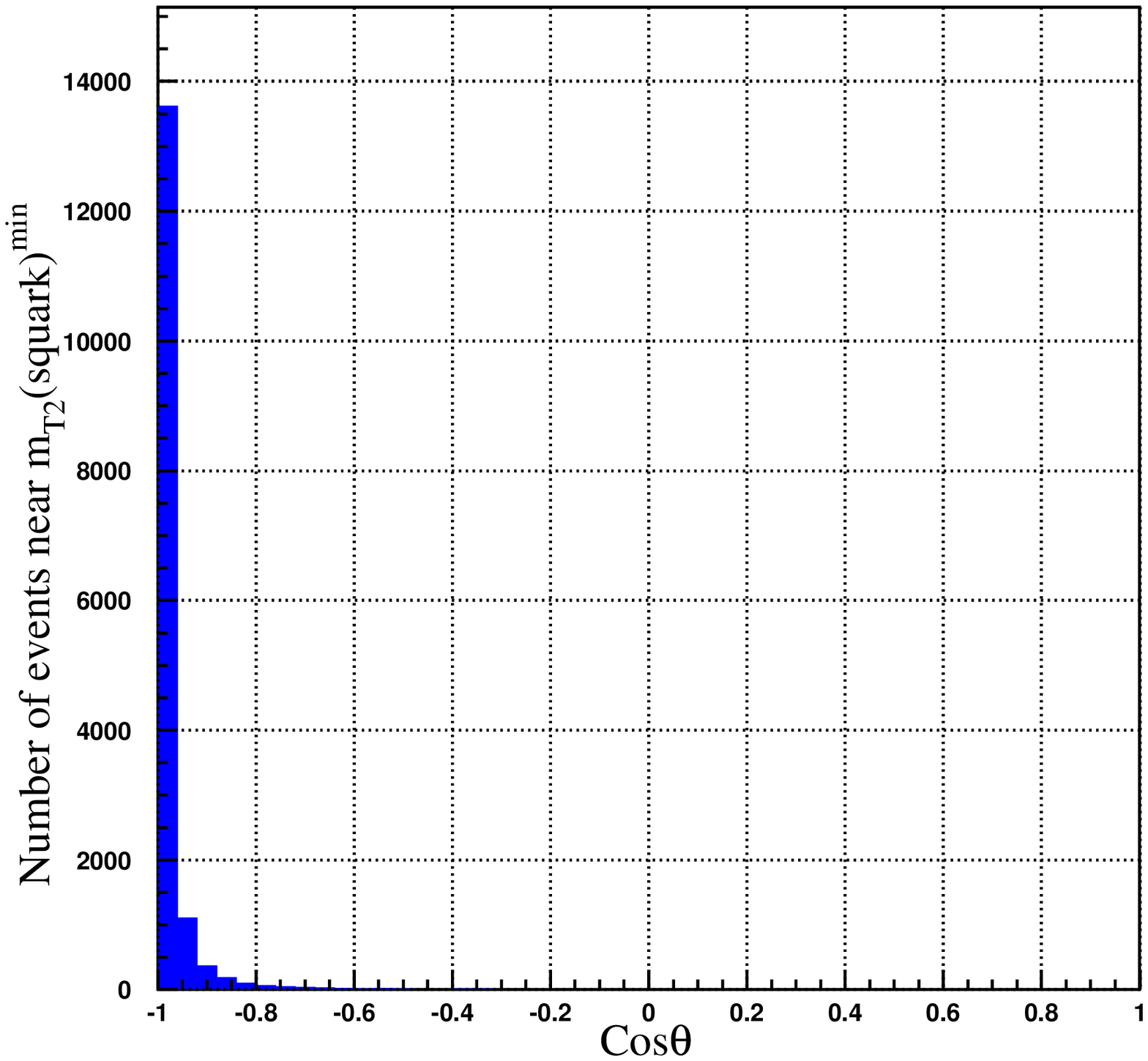,width=7cm,height=7cm}
\epsfig{figure=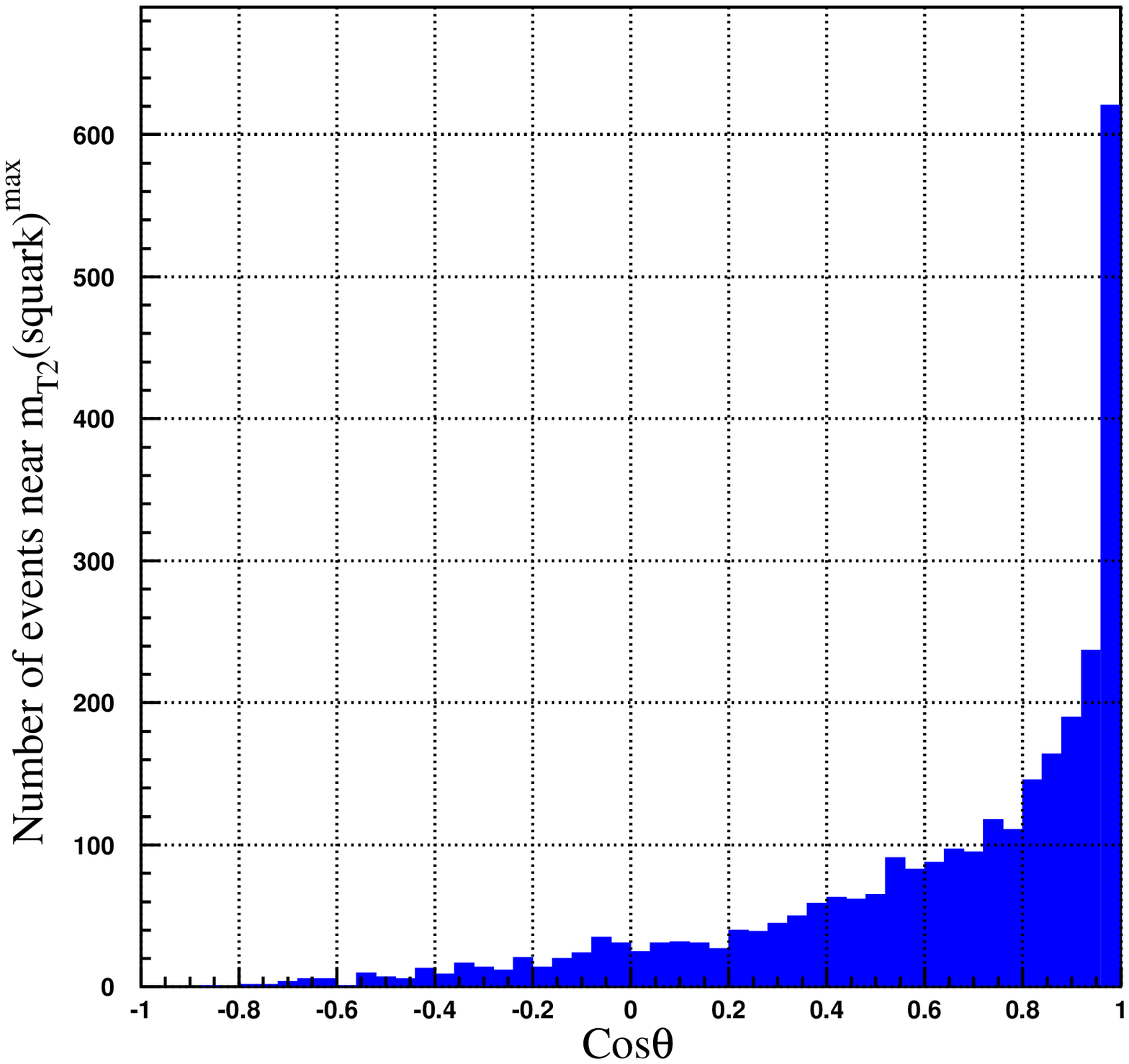,width=7cm,height=7cm}
\end{center}
\caption{Event distribution along $\cos\theta$ for the events near
(a) the lower edge and (b) the upper edge region of the Fig.
\ref{fig:mirage}(a). } \label{fig:mirage-theta}
\end{figure}

Fig. \ref{fig:mirage-theta} shows the distribution of the opening
angle $\theta$ between the two visible quark momenta for the events
near (a) the lower edge and (b) the upper edge of Fig.
\ref{fig:mirage}(a). As anticipated, the events near the lower edge
and the upper edge  give a distribution with a peak at $\cos
\theta=-1$ and $\cos \theta=1$, respectively. The rather broad peak
in Fig. \ref{fig:mirage-theta}(b) is due to the non-zero transverse
momentum of the each squark, which makes the two quarks from the
squark-pair decay less aligned in general.

%

\subsection{Gluino $m_{T2}$} \label{sec:gluino_mt2}

Let us now examine  a more complicate process in which each of the
pair-produced mother particles decays into one invisible LSP and
{\it more than one} visible particles, for which $m_{T2}^{\rm
max}(m_\chi)$ shows a kink structure at
$m_\chi=m_{\tilde{\chi}_1^0}$ as was noticed first in \cite{mt203}
and discussed in the previous section in a generic context. As a
specific example, we consider the process in which a pair of gluinos
are produced in proton-proton collision and each gluino decays into
two quarks and one LSP:
\begin{eqnarray}
pp \rightarrow {\tilde g}{\tilde g} \rightarrow qq\tilde\chi_1^0
qq\tilde\chi_1^0, \label{gluino_decay}
\end{eqnarray}
where again $q$ denotes the 1st or 2nd generation quark. Depending
upon whether squarks are heavier or lighter than gluino,  the gluino
decay $\tilde g \rightarrow qq\tilde\chi_1^0$ occurs through a
three-body decay induced by an exchange of off-shell squark or two
body cascade decay with intermediate on-shell squark. As noticed in
the previous section, these two cases have  a different value of
$m_{vis}^{\rm max}=m_{qq}^{\rm max}$, thereby the resulting
$m_{T2}^{\rm max}(m_\chi)$ has a different functional form in the
range $m_\chi
> m_{\tilde{\chi}^0_1}$.

For the process (\ref{gluino_decay}),
the unconstrained minima of the transverse masses of the decay
product of each gluino, i.e.
 $\big(m_T^{(i)}\big)_{\rm min}=m_{vis}^{(i)}+m_\chi$,
are generically different from each other, thereby both of the
balanced and unbalanced $m_{T2}$ solutions can appear.

If the squark mass $m_{\tilde q}$ is heavier than the gluino mass
$m_{\tilde g}$, gluino will undergo the three body decay $\tilde g
\rightarrow qq\tilde\chi_1^0$ through an exchange of off-shell
squark. In this case, for $m_q=0$, the total invariant
mass\footnote{One might consider the total transverse mass of the
visible part which has the same range.} of the visible part of each
gluino decay is in the following range:
\begin{eqnarray}
0 \,\leq\, m_{vis}^{(1)},\,m_{vis}^{(2)}\, \leq \,m_{\tilde
g}-m_{\tilde\chi_1^0}.
\end{eqnarray}

As discussed in section 2, in order to obtain the maximum of gluino
$m_{T2}$,  we can limit ourselves to the situation that the two
gluinos are produced at rest and all decay products are moving on
the transverse plane.
In such case, the transverse momentum and energy of the visible part
are given by
\begin{eqnarray}
|{\bold p}_T^{vis(i)}|&=&\frac{\sqrt{\big((m_{\tilde
g}+m_{vis}^{(i)})^2-m_{\tilde\chi_1^0}^2\big) \big((m_{\tilde
g}-m_{vis}^{(i)})^2-m_{\tilde\chi_1^0}^2\big)}}{2 m_{\tilde
g}} \quad (i=1,2), \nonumber \\
E_T^{vis(i)}&=&{m_{\tilde
g}^2-m_{\tilde\chi_1^0}^2+(m_{vis}^{(i)})^2 \over 2 m_{\tilde g}}.
 \label{gmt2-pt}
\end{eqnarray}
Then, the balanced $m_{T2}$ solution (\ref{balanced}) is
unambiguously fixed by the four variables: $m_{vis}^{(1)}$,
$m_{vis}^{(2)}$, $\theta=$ the angle between ${\bold p}_T^{vis(1)}$
and ${\bold p}_T^{vis(2)}$, and the trial LSP mass $m_\chi$.

For given values of  $m_{vis}^{(i)}$ and $m_\chi$, the maximum of
the balanced $m_{T2}$ occurs at $\theta=0$ (see the appendix B for a
detailed proof.)
Since we are mainly interested in $m_{T2}^{\rm max}$, we will focus
on the configuration with $\theta=0$ in the following. From
(\ref{balanced}), we obtain
\begin{eqnarray}
&& m_{T2}^2|_{\theta=0} \, =\, m_\chi^2 + \tilde{A} \nonumber \\
&&\quad +\,\sqrt{\left(1+ {4 m_\chi^2 \over 2
\tilde{A}-\left(m_{vis}^{(1)}\right)^2-\left(m_{vis}^{(2)}\right)^2
}\right) \left(\tilde{A}^2 -\Big(m_{vis}^{(1)}m_{vis}^{(2)}\Big)^2
\right)},
 \label{gmt2-bal}
\end{eqnarray}
where
\begin{eqnarray}
\tilde{A}\equiv A|_{\theta=0}=E_T^{vis(1)} E_T^{vis(2)} + |{\bold
p}_T^{vis(1)}||{\bold p}_T^{vis(2)}|,
\end{eqnarray}
with $|{\bold p}_T^{vis(i)}|$ and $E_T^{vis(i)}$ given by
(\ref{gmt2-pt}). If we further restrict the momentum configurations
to satisfy
\begin{eqnarray}
m_{vis}^{(1)}=m_{vis}^{(2)}\equiv m_{vis}, \end{eqnarray} the
balanced $m_{T2}$ solution is simplified as
\begin{eqnarray}
m_{T2}|_{\theta=0}&=& {m_{\tilde g}^2-m_{\tilde\chi_1^0}^2+m_{vis}^2
\over 2 m_{\tilde g}}
\nonumber \\
&+&\sqrt{{\big((m_{\tilde
g}+m_{vis})^2-m_{\tilde\chi_1^0}^2\big)\big((m_{\tilde
g}-m_{vis})^2-m_{\tilde\chi_1^0}^2\big)\over 4m^2_{\tilde g}} +
m_\chi^2}. \label{gmt2}
\end{eqnarray}
Then, from (\ref{mt2max}), we obtain
\begin{eqnarray}
\label{gmt2-leq} m_{T2}^{\rm max}(m_\chi) &=& {m_{\tilde
g}^2-m_{\tilde\chi_1^0}^2 \over 2 m_{\tilde g}}
+\sqrt{\left({m_{\tilde g}^2-m_{\tilde\chi_1^0}^2 \over 2 m_{\tilde
g}}\right)^2 +m_\chi^2}\quad  {\rm if}~~m_\chi < m_{\tilde\chi_1^0},
\\
m_{T2}^{\rm max}(m_\chi) &=& \left(m_{\tilde
g}-m_{\tilde\chi_1^0}\right)+m_\chi, ~~~~~~~~~~~~~~~~~~~~~~ \quad
{\rm if}~~m_\chi > m_{\tilde\chi_1^0}. \label{gmt2-geq}
\end{eqnarray}

\begin{figure}[ht!]
\begin{center}
\epsfig{figure=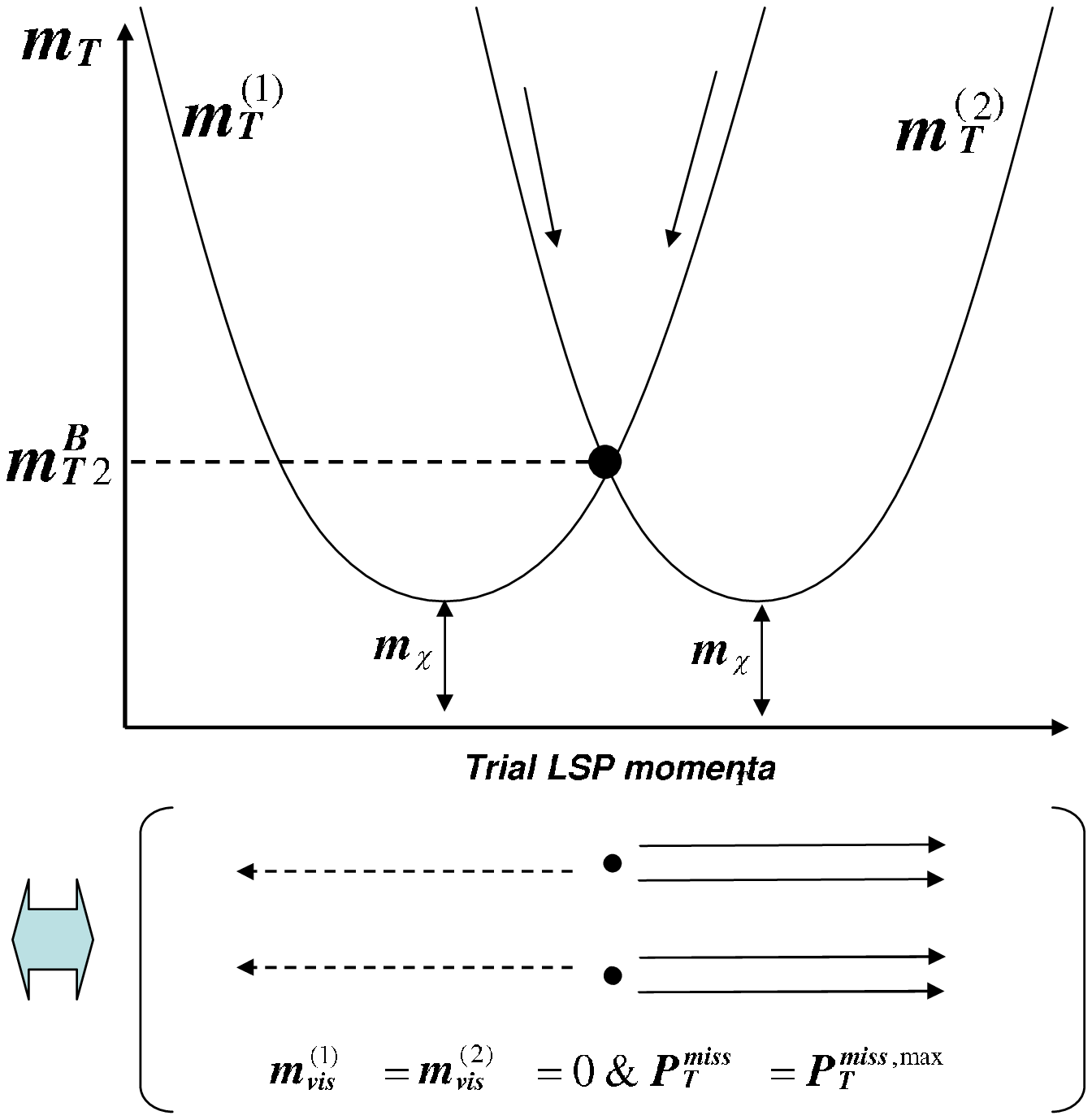,width=7cm,height=7cm}
\epsfig{figure=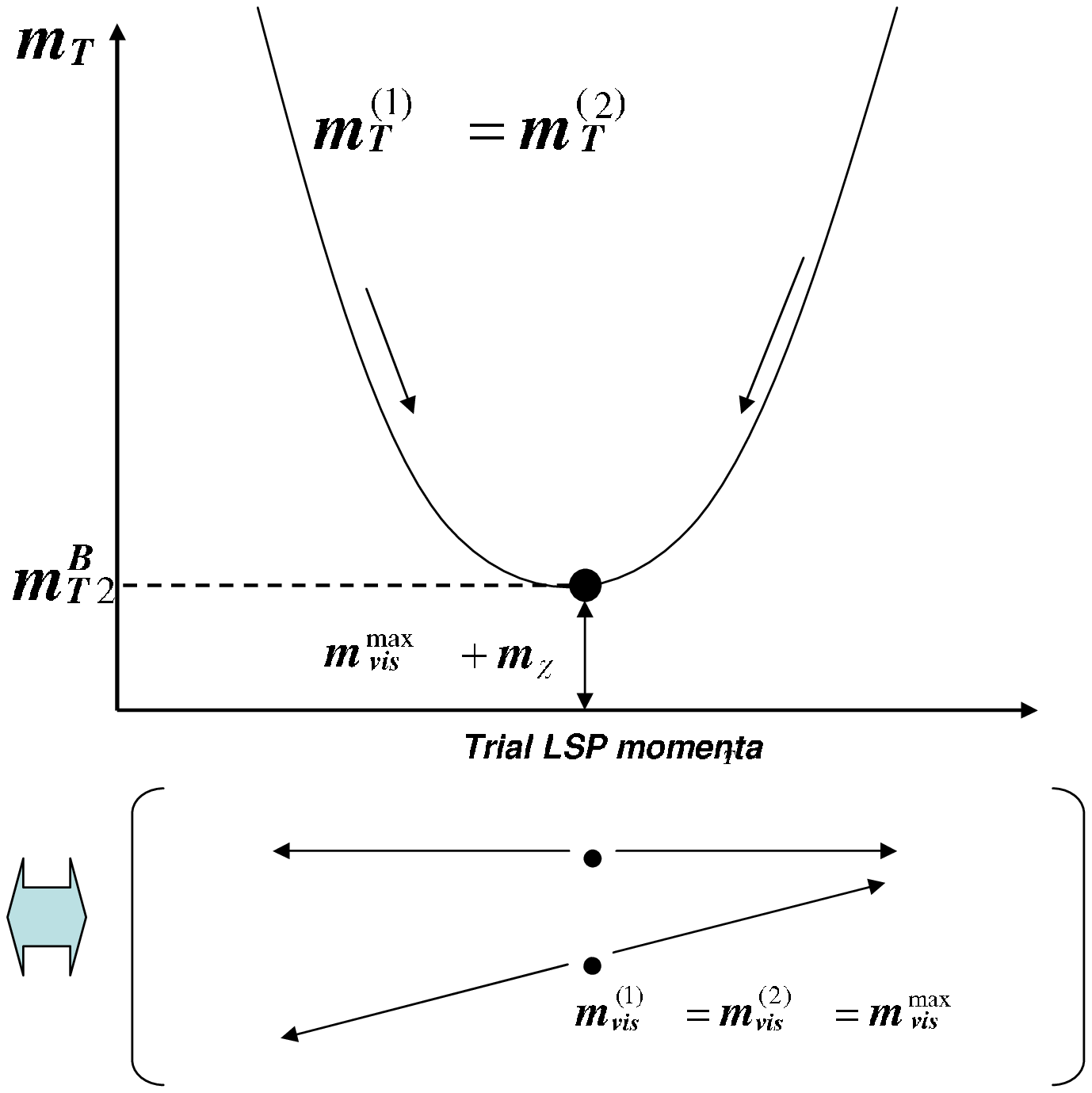,width=7cm,height=7cm}
\end{center}
\caption{Extreme momentum configuration providing $m_{T2}^{\rm max}$
as a balanced solution for (a) $m_\chi<m_{\tilde\chi_1^0}$ and (b)
$m_\chi>m_{\tilde\chi_1^0}$. } \label{fig:mm}
\end{figure}

Fig. \ref{fig:mm}(a) shows a momentum configuration providing the
$m_{T2}^{\rm max}$ of (\ref{gmt2-leq}). In this configuration, two
gluinos are produced at rest, and each gluino subsequently decays
into two quarks moving in the same direction (i.e.
$m_{vis}^{(1)}=m_{vis}^{(2)}=0$) and one LSP moving in the opposite
direction. Furthermore, two sets of gluino decay products are
parallel to each other (i.e. $\theta=0$) and all of them are on the
transverse plane with respect to the proton beam direction. This
configuration is the first example of extreme momentum configuration
considered in Ref. \cite{mt203}. Although this corresponds to the
simplest configuration providing the $m_{T2}^{\rm max}$ of
(\ref{gmt2-leq}), it might not be useful for constructing
$m_{T2}^{\rm max}$ from real collider data as all quarks are moving
in the same direction, so that they can not be identified as
separate particles due to the finite jet resolution.

This difficulty of jet resolution can be partly avoided by the
back-to-back transverse boost of the above extreme configuration,
giving the same value of $m_{T2}^{\rm max} (m_\chi)$. In the
back-to-back boosted configurations, the two di-quark systems are
{\it not} moving in the same direction in general, so that can be
distinguished from each other in real collider event. However, the
two quarks in each di-quark system are still aligned to each other.
In real collider data analysis, two aligned quarks cannot be
identified as separate jets with realistic jet reconstruction, which
will eliminate the events which involve the quarks moving in the
same direction. As the true maximum of $m_{T2}$ comes from such
momentum configuration, any
realistic jet reconstruction will cause a systematic shift of
$m_{T2}^{\rm max}$ to a lower value when one tries to construct
$m_{T2}^{\rm max}$ from real collider data.  Our analytic expression
(\ref{balanced}) for the balanced $m_{T2}$ solution provides
information on how sensitive $m_{T2}$ is to the angular separation
of the involved quarks, with which one can estimate the uncertainty
of $m_{T2}^{\rm max}$ caused by the jet resolution cut:
\begin{eqnarray}
\frac{\Delta m_{T2}^{\mathrm{max}}}{m_{T2}^{\mathrm{max}}}&\approx &
-\frac{1}{8}\frac{m_{\widetilde{\chi}_{1}^{0}}^{2}}{m_{\widetilde{g}}^{2}}%
\left(1-\frac{m_{\widetilde{\chi}_{1}^{0}}^{2}}{m_{\widetilde{g}}^{2}}\right)\big(\Delta
R\big)^{2}\, \lesssim\, {\cal O}(1) \, \%,
\end{eqnarray}
where $\Delta R\equiv\sqrt{\Delta\phi^{2}+\Delta\eta^{2}}\sim 0.5$
represents a separation of two quarks in azimuthal angle and
pseudorapidity plane.
This indicates that the systematic shift of $m_{T2}^{\rm max}$ due
to the finite jet resolution is negligible, which we have confirmed
by an explicit  Monte Carlo analysis.

The momentum configuration of Fig. \ref{fig:mm}(b) provides the
$m_{T2}^{\rm max}$ of (\ref{gmt2-geq}), which was considered in
\cite{mt203}  as the second example of extreme momentum
configuration. Here, gluinos are pair produced at rest, the two
quarks from each gluino are back to back to each other
($m_{vis}^{(1)}=m_{vis}^{(2)}=m_{\tilde g}-m_{\tilde\chi_1^0}$),
while the LSP is at rest. In this case, the angle $\theta$ is not
well defined because ${\bold p}_T^{vis(1)}={\bold p}_T^{vis(1)}=0$.
Also
 ${\bold p}_T^{miss}=0$ which is true even after a back-to-back
boost of the system. Such momentum configuration will be useless
when one constructs $m_{T2}^{\rm max}$ from collider data as one
typically uses an event selection cut imposing a lower bound on
$|{\bold p}_T^{miss}|$. However, as we will see shortly, there exist
momentum configurations yielding (\ref{gmt2-geq}) while having a
sizable $|{\bold p}_T^{miss}|$, so that the $m_{T2}^{\rm max}$ of
(\ref{gmt2-geq}) can be constructed from  collider data even under a
proper cut on $|{\bold p}_T^{miss}|$.

\begin{figure}[ht!]
\begin{center}
\epsfig{figure=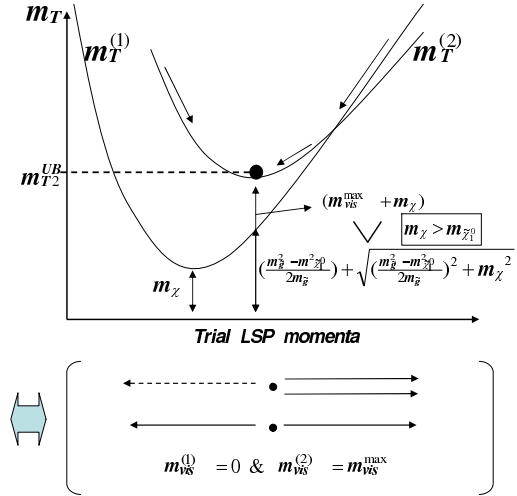,width=7cm,height=7cm}
\end{center}
\caption{Extreme momentum configuration providing $m_{T2}^{\rm max}$
as an unbalanced solution for $m_\chi>m_{\tilde\chi_1^0}$.}
\label{fig:xx}
\end{figure}

A momentum configuration, which provides the $m_{T2}^{\rm max}$ of
(\ref{gmt2-geq}) with a sizable $|{\bold p}_T^{miss}|$, is shown in
Fig.\ref{fig:xx}. In this configuration, two gluinos are produced at
rest. The first gluino produces a di-quark system with
$m_{vis}^{(1)}=0$ and one LSP, while the second gluino produces a
back-to-back di-quark system with $m_{vis}^{(2)}=m_{\tilde
g}-m_{\tilde\chi_1^0}$ and one LSP at rest. For the second gluino
decay set, the visible momentum ${\bold p}_T^{vis(2)}=0$. Thus,
unconstrained minimum of the second gluino transverse mass,
$m_T^{(2)}=(m_{\tilde g}-m_{\tilde\chi_1^0})+m_\chi$, occurs when
trial LSP momentum ${\bold p}_T^{\chi(2)}=0$. On the other hand, the
first gluino decay product has $|{\bold p}_T^{vis(1)}|=(m_{\tilde
g}^2-m_{\tilde\chi_1^0}^2)/2m_{\tilde g}$ and ${\bold
p}_T^{\chi(1)}=-{\bold p}_T^{vis(1)}$, for ${\bold
p}_T^{\chi(2)}=0$. Then, the corresponding transverse mass of the
first gluino decay is given by
\begin{eqnarray}
m_T^{(1)}={m_{\tilde g}^2-m_{\tilde\chi_1^0}^2 \over 2m_{\tilde g}}
+\sqrt{\left({m_{\tilde g}^2-m_{\tilde\chi_1^0}^2\over2m_{\tilde
g}}\right)^2+m_\chi^2}.
\end{eqnarray}
Therefore, $m_T^{(2)} > m_T^{(1)}$ if $m_\chi>m_{\tilde\chi_1^0}$,
though $m_T^{(2)}$ is at the unconstrained minimum value, so that we
have unbalanced $m_{T2}$ solution, i.e. $m_{T2}=(m_{\tilde
g}-m_{\tilde\chi_1^0})+m_\chi$ for this momentum configuration. The
same unbalanced $m_{T2}$ solution is obtained for the momentum
configuration with other values of $m_{vis}^{(1)}$  because those
cases also give $m_T^{(2)}\geq m_T^{(1)}$ when $m_T^{(2)}$ is at the
unconstrained minimum. Such momentum configurations and the
back-to-back transverse boosted ones would have a sizable $|{\bold
p}_T^{miss}|$,  so can be used to determine $m_{T2}^{\rm max}$ from
real collider data.

To summarize the extremal features of $m_{T2}$ for the decay of
gluino pair when $m_{\tilde{q}}>m_{\tilde{g}}$, the maximum of
$m_{T2}$ over all events is given by
\begin{eqnarray}
m_{T2}^{\rm max}(m_\chi) &=& {m_{\tilde g}^2-m_{\tilde\chi_1^0}^2
\over 2 m_{\tilde g}} +\sqrt{\left({m_{\tilde
g}^2-m_{\tilde\chi_1^0}^2 \over 2 m_{\tilde g}}\right)^2
+m_\chi^2}\quad  {\rm if}~~m_\chi < m_{\tilde\chi_1^0}, \nonumber
\\
m_{T2}^{\rm max}(m_\chi) &=& \left(m_{\tilde
g}-m_{\tilde\chi_1^0}\right)+m_\chi \quad {\rm if}~~m_\chi >
m_{\tilde\chi_1^0} \label{result1}
\end{eqnarray}
as obtained in (\ref{multivis1}) and (\ref{multivis2}) in more
generic context.
Thus, there is a level crossing of $m_{T2}^{\rm max}$ at $m_\chi =
m_{\tilde{\chi}^0_1}$, yielding  a kink structure  as shown in Fig.
\ref{fig:mt2_heavy_lightsquark}(a). If such  $m_{T2}^{\rm
max}$-curve can be constructed from collider data, which will be
examined in the next subsection,  this kink structure will enable us
to determine the true LSP mass $m_{\tilde{\chi}^0_1}$ and the gluino
mass $m_{\tilde{g}}= m_{T2}^{\rm max}(m_\chi=m_{\tilde{\chi}^0_1})$
{\it simultaneously}.



\begin{figure}[ht!]
\begin{center}
\epsfig{figure=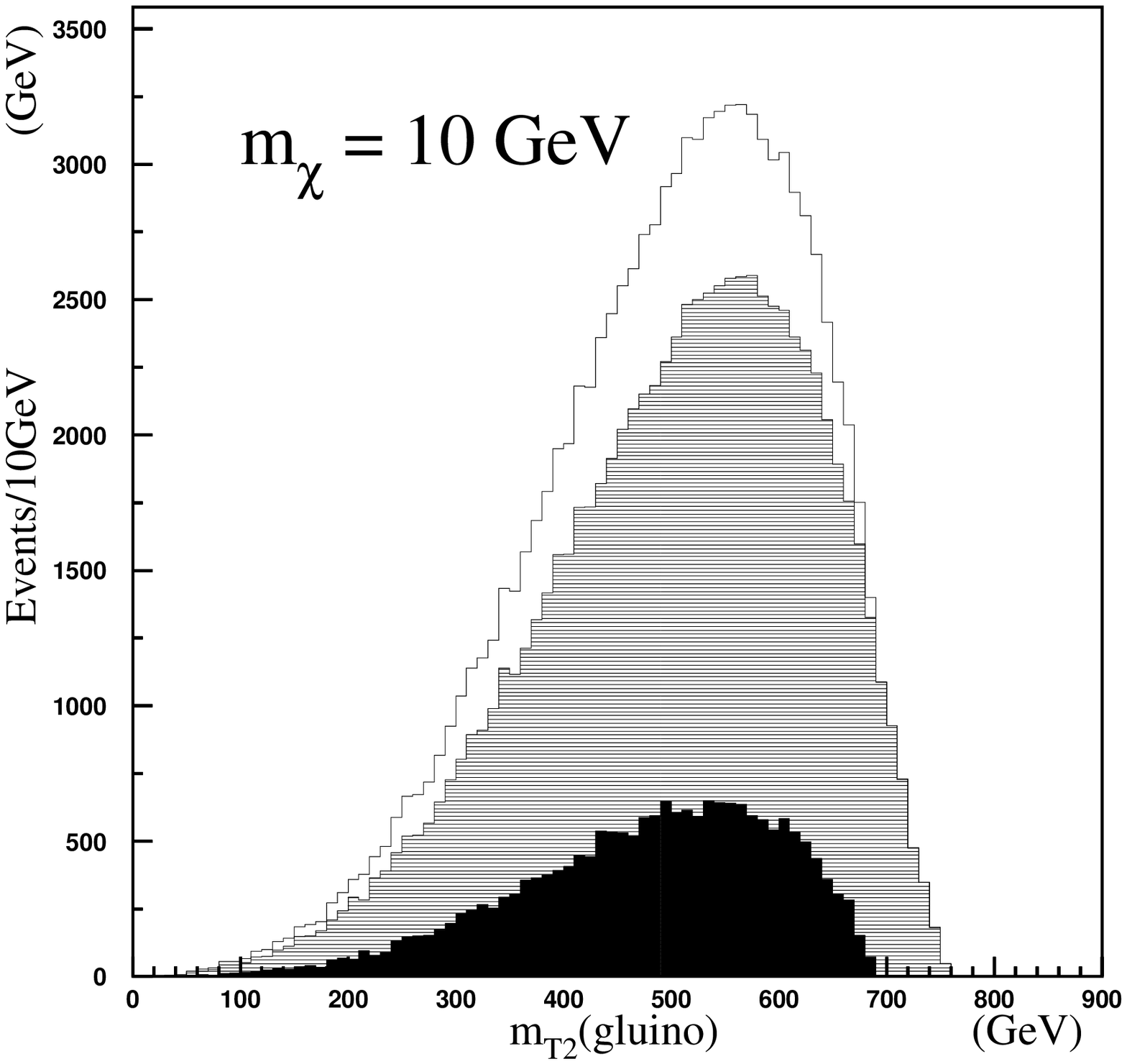,width=7cm,height=7cm}
\epsfig{figure=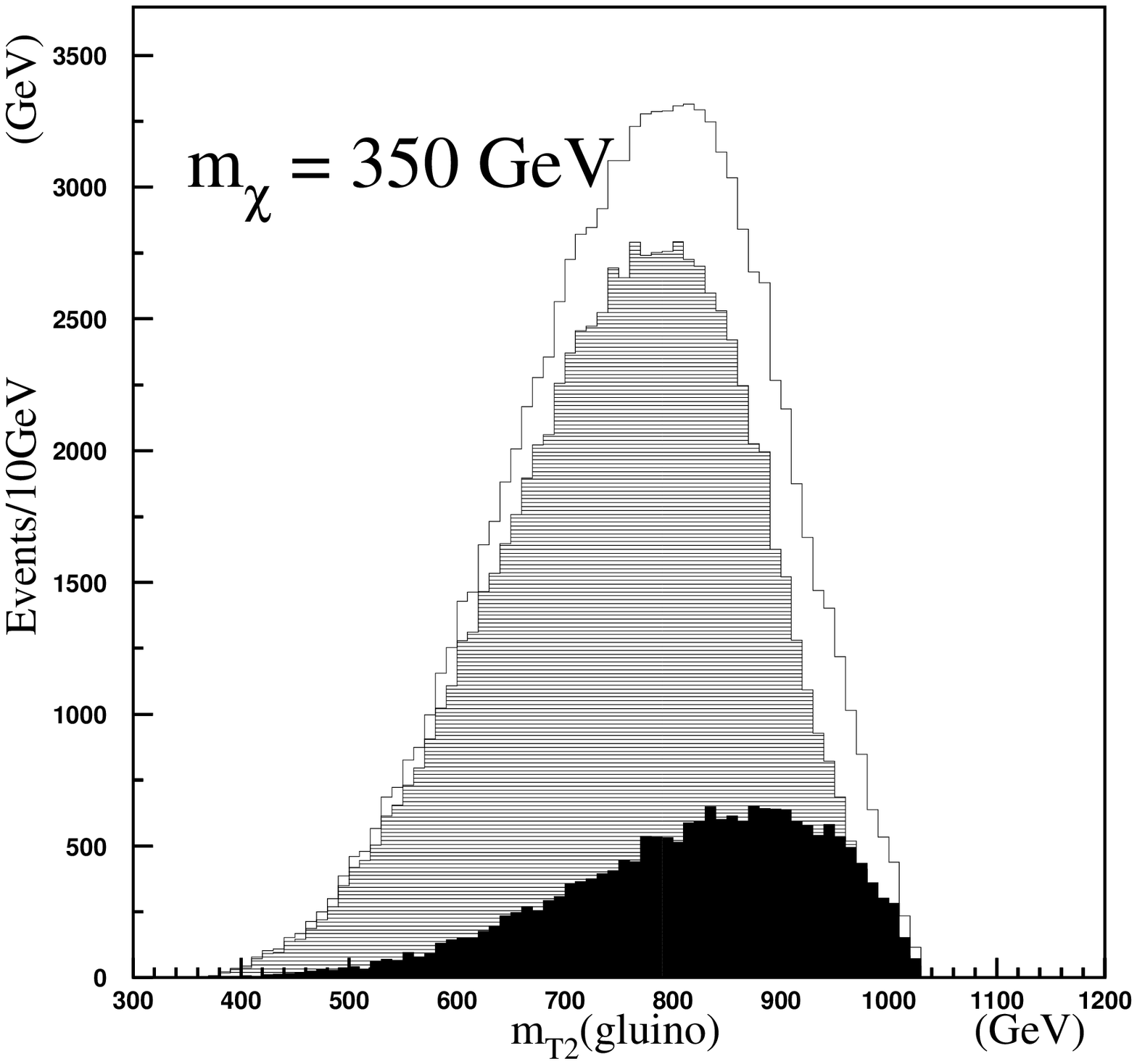,width=7cm,height=7cm}
\end{center}
\caption{$m_{T2}$ distribution with (a) $m_\chi=10$ GeV and (b)
$m_\chi=350$ GeV for the AMSB parameter point (\ref{gmt2-mass}). }
\label{fig:gmt2-parton}
\end{figure}

\begin{figure}[ht!]
\begin{center}
\epsfig{figure=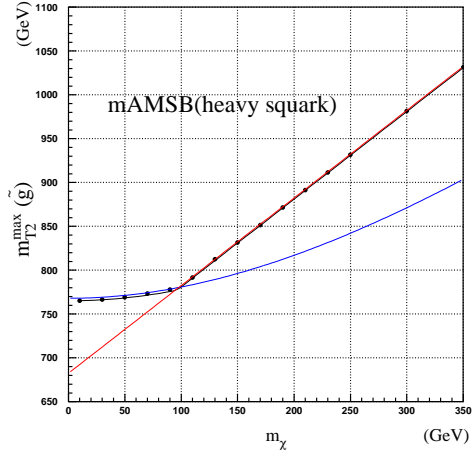,width=7cm,height=7cm}
\end{center}
\caption{$m_{T2}^{\rm max}$ as a function of the trial LSP mass
$m_\chi$ for the AMSB parameter point (\ref{gmt2-mass}). }
\label{fig:gmt2-curve}
\end{figure}

To see explicitly the extremal features of gluino $m_{T2}$, a Monte
Carlo event sample of the signal $pp \rightarrow \tilde g\tilde
g\rightarrow qq{\tilde\chi_1^0} qq{\tilde\chi_1^0}$ has been
generated in partonic-level, for a SUSY parameter point in a minimal
anomaly mediated SUSY-breaking (AMSB) scenario \cite{amsb}, which
gives
\begin{eqnarray}
m_{\tilde g} = 780~{\rm GeV},~~m_{\tilde\chi_1^0}=98~{\rm GeV},
\label{gmt2-mass}
\end{eqnarray} with a few TeV sfermion masses. The $m_{T2}$ values
for the event sample were then calculated. Fig.\ref{fig:gmt2-parton}
(a) and (b) show the resulting $m_{T2}$ distributions for trial LSP
mass $m_\chi=10$ GeV and 350 GeV, respectively. On the figures,
hatched histogram corresponds to the balanced $m_{T2}$ values, while
black histogram to the unbalanced ones. As anticipated, one can
notice that for $m_\chi=10$ GeV, which is smaller than true LSP
mass, the endpoint of the $m_{T2}$ distribution is determined by the
balanced $m_{T2}$ solutions, while both balanced and unbalanced
$m_{T2}$ solution contribute to the endpoint region for $m_\chi=350$
GeV, which is larger than true LSP mass. Finally, $m_{T2}^{\rm max}$
as a function of the trial LSP mass $m_\chi$ is shown in Fig.
\ref{fig:gmt2-curve}. Here, the blue and red curves represent the
analytic formula (\ref{result1}), while the black dots are obtained
from the Monte Carlo data, which fit very well the analytic curves.

Let us now consider the case of lighter squark, $m_{\tilde q}<
m_{\tilde g}$, for which the following cascade decay is open:
\begin{eqnarray}{\tilde g} \rightarrow q{\tilde q} \rightarrow
qq\tilde\chi_1^0,\end{eqnarray} where the squark in the second stage
is on mass-shell. The main difference between this two body cascade
decay and the three body decay is that the total invariant mass of
the visible part takes the range:
\begin{eqnarray}
0 \,\leq\, m_{vis}^{(1)}, \,m_{vis}^{(2)} \,\leq\, \sqrt{{(m_{\tilde
g}^2 - m_{\tilde q}^2) (m_{\tilde q}^2 - m_{\tilde\chi_1^0}^2) \over
m_{\tilde q}^2}}.
\end{eqnarray}
As $m_{vis}^{\rm max}$ has a smaller value than the three body decay
case,  while $m_{vis}^{\rm min}$ is same, the kink structure is
weakened as was anticipated in (\ref{cusp1}). The maximum of
$m_{T2}$ for $m_\chi < m_{\tilde\chi_1^0}$  takes the same form as
the case of heavier squarks, while it is changed to a different form
for $m_\chi
> m_{\tilde\chi_1^0}$:
\begin{eqnarray}
m_{T2}^{\rm max}(m_\chi) &=& {m_{\tilde g}^2-m_{\tilde\chi_1^0}^2
\over 2 m_{\tilde g}} +\sqrt{\left({m_{\tilde
g}^2-m_{\tilde\chi_1^0}^2 \over 2 m_{\tilde g}}\right)^2
+m_\chi^2}\quad {\rm if}~~m_\chi < m_{\tilde\chi_1^0}, \nonumber \\
m^{\rm max}_{T2} (m_\chi) &=& \left({m_{\tilde g} \over 2}
(1-{m_{\tilde q}^2 \over m_{\tilde g}^2}) +{m_{\tilde g} \over 2}
(1-{m_{\tilde\chi_1^0}^2 \over m_{\tilde q}^2})\right)\nonumber \\
&+&\sqrt{\left({m_{\tilde g} \over 2} (1-{m_{\tilde q}^2 \over
m_{\tilde g}^2}) -{m_{\tilde g} \over 2} (1-{m_{\tilde\chi_1^0}^2
\over m_{\tilde q}^2})\right)^2+m_\chi^2} \quad {\rm if}~~m_\chi >
m_{\tilde\chi_1^0}\label{result2}
\end{eqnarray}
as obtained in (\ref{multivis3}). The $m_{T2}^{\rm max}$-curve for
lighter squarks is
depicted  in Fig. \ref{fig:mt2_heavy_lightsquark}(b), which shows
again a kink structure at $m_\chi = m_{\tilde{\chi}^0_1}$, although
milder than the case of heavier squarks. Note that the $m_{T2}^{\rm
max}$-curve for the range $m_\chi > m_{\tilde{\chi}^0_1}$ depends on
the squark mass also, so it can determine the gluino mass, the LSP
mass and the squark mass altogether.


\subsection{Construction of gluino $m_{T2}$ from collider data}
\label{sec:experiment}

In order to check the experimental feasibility of measuring
superparticle masses using the kink structure of the gluino
$m_{T2}^{\rm max}$, we have generated Monte Carlo event samples of
the SUSY signals at LHC by PYTHIA \cite{pythia} for several SUSY
breaking schemes yielding different patterns of superparticle
spectra.
We have also generated the SM backgrounds such as $t \bar{t}$, $W/Z
+ \mbox{jet}$, $WW/WZ/ZZ$ and QCD events, with less equivalent
luminosity, in five logarithmic $p_T$ bins for 50 GeV $< p_T <$ 4000
GeV. The SM backgrounds have been also generated by PYTHIA.
The generated events have been further processed with a modified
version of the fast detector simulation program PGS \cite{pgs},
which approximates an ATLAS or CMS-like detector with reasonable
efficiencies and fake rates.

For each event, the four leading jets are used to calculate the
gluino $m_{T2}$. For a convenience of numerical analysis, we
considered $m_{T2}$ defined in terms of the transverse visible mass
$m_T^{vis(i)}$, rather than in terms of the invariant visible mass
$m_{vis}^{(i)}$, which gives the same value of $m_{T2}^{\rm max}$ as
remarked in section 2. Note that $m_T^{vis(i)}=m_{vis}^{(i)}$ for
the extreme momentum configurations giving the maximal value of
$m_{T2}$. The four jets  are divided in two groups of dijets as
follows \cite{hemi}. The highest momentum jet and the other jet
which has the largest $\left | p_{jet} \right | \Delta R$ with
respect to the leading jet are chosen as two `seed' jets for the
division. Here, $p_{jet}$ is the jet momentum and $\Delta R \equiv
\sqrt{\Delta \phi^2 + \Delta \eta^2}$ denotes the jet separation in
azimuthal angle and pseudorapidity plane. Each of the remaining two
jets is associated to the seed jet making a smaller opening angle.
Then, each of the jet pairs constructed in this way is considered to
be originating from the same mother particle (gluino). If this
procedure fails to choose two groups of jet pairs, we discarded the
event.

Because the functional form of the gluino $m_{T2}^{\rm max} ( m_\chi
)$ depends upon whether the 1st and 2nd generations of squarks are
heavier than gluino or not, we consider those two cases separately.
Let us first consider the case of heavier squarks. Superparticle
spectrum with $m_{\tilde{q}}
> m_{\tilde{g}}$ can arise from various SUSY breaking schemes,
for which the gluino $m_{T2}$ takes the form of Fig.
\ref{fig:mt2_heavy_lightsquark}(a). For simplicity, here we consider
only the case that the 1st and 2nd generations of squarks are
significantly heavier than 1 TeV, e.g. $m_{\tilde{q}}\sim 4$ TeV, so
that squarks are {\it not} copiously produced at LHC\footnote{If
squarks are heavier than gluino, but still light enough to be
copiously produced at LHC, the gluino $m_{T2}$ is not a proper
observable to determine the gluino and LSP masses since the
gluino-pair events $\tilde{g}\tilde{g}\rightarrow
qq\tilde{\chi}_1^0qq\tilde{\chi}_1^0$ are severely screened by the
squark pair events $\tilde{q}\tilde{q}\rightarrow
q\tilde{g}q\tilde{g}\rightarrow
qqq\tilde{\chi}_1^0qqq\tilde{\chi}_1^0$. In such case, one can
construct the squark $m_{T2}$ for the squark pair events with six
quarks, which would show a behavior similar to the gluino
$m_{T2}^{\rm max}$ in the case of lighter squark, from which the
squark, gluino, and LSP masses can be determined altogether.}, while
gluino is light enough to be copiously produced, e.g.
$m_{\tilde{g}}< 1$ TeV.
As an specific example of such superparticle spectrum, here we
consider a parameter point of anomaly mediation scenario (AMSB) with
heavy sfermion masses, in which the gluino, LSP and (the 1st and 2nd
generation) squark masses are given by\footnote{We also assume
$\tan\beta=10$ and the Higgsino mass parameter $\mu >0$.}
\begin{eqnarray}
\mbox{AMSB with heavy sfermions}:\, m_{\tilde{g}}=780 \, {\rm GeV},
\quad m_{\tilde{\chi}_1^0}=98 \, {\rm GeV}, \quad m_{\tilde{q}}=4 \,
{\rm TeV}.
\label{benchmark}\end{eqnarray}
For the AMSB point, the production cross section of gluino pair
$\sigma({\tilde g}{\tilde g})\sim 1.1 {\rm pb}$. The branching
ratios of gluino decay are as follows: $B(\tilde g\rightarrow
{\tilde\chi_1^0 qq}) \sim 32 \%$, $B(\tilde g\rightarrow
{\tilde\chi_2^0 qq}) \sim 3 \%$, $B(\tilde g\rightarrow
{\tilde\chi_1^{\pm} qq'}) \sim 64 \%$. Thus, gluino mostly decays
into lighter chargino (or LSP) plus two quarks. Being wino-like, the
LSP and the lighter chargino are almost degenerate in mass. The
chargino decay $\tilde\chi_1^{\pm} \rightarrow {\tilde\chi_1^0}
l^{\pm}\nu$ produces very soft leptons, which cannot be detected at
LHC. In this circumstance, both gluino decays ${\tilde g}
\rightarrow \tilde\chi_1^{\pm} qq'$ and $\tilde g\rightarrow
{\tilde\chi_1^0}qq$ can be considered as `signals' we are looking
for, and the contamination from the small number of $\tilde
g\rightarrow {\tilde\chi_2^0} qq$ decay is expected not to be
significant. In this work, we assume integrated luminosity of 300
$fb^{-1}$ for the AMSB point.


%
\begin{figure}[ht!]
\begin{center}
\epsfig{file=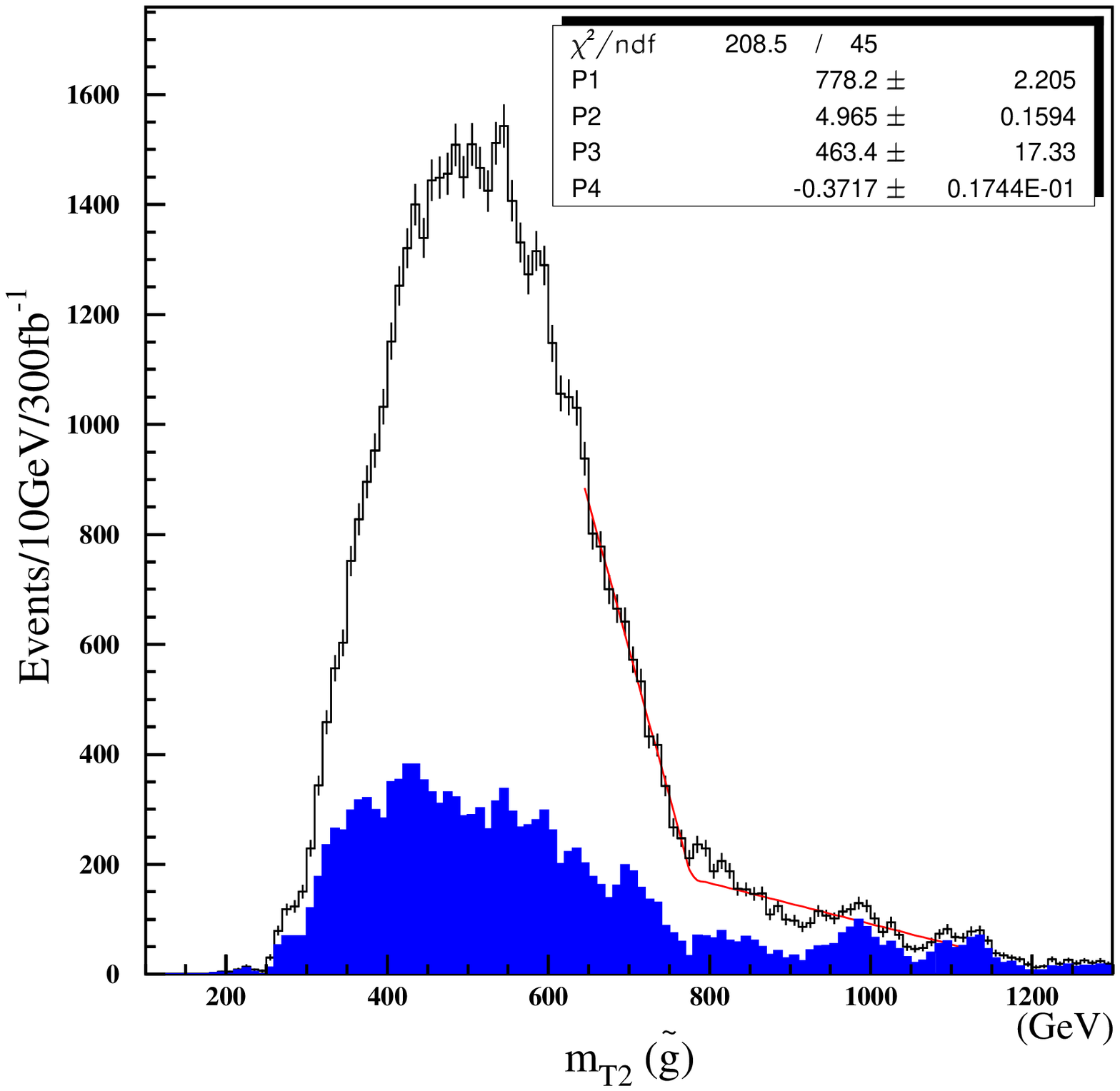, width=6cm, height=6cm}
\epsfig{file=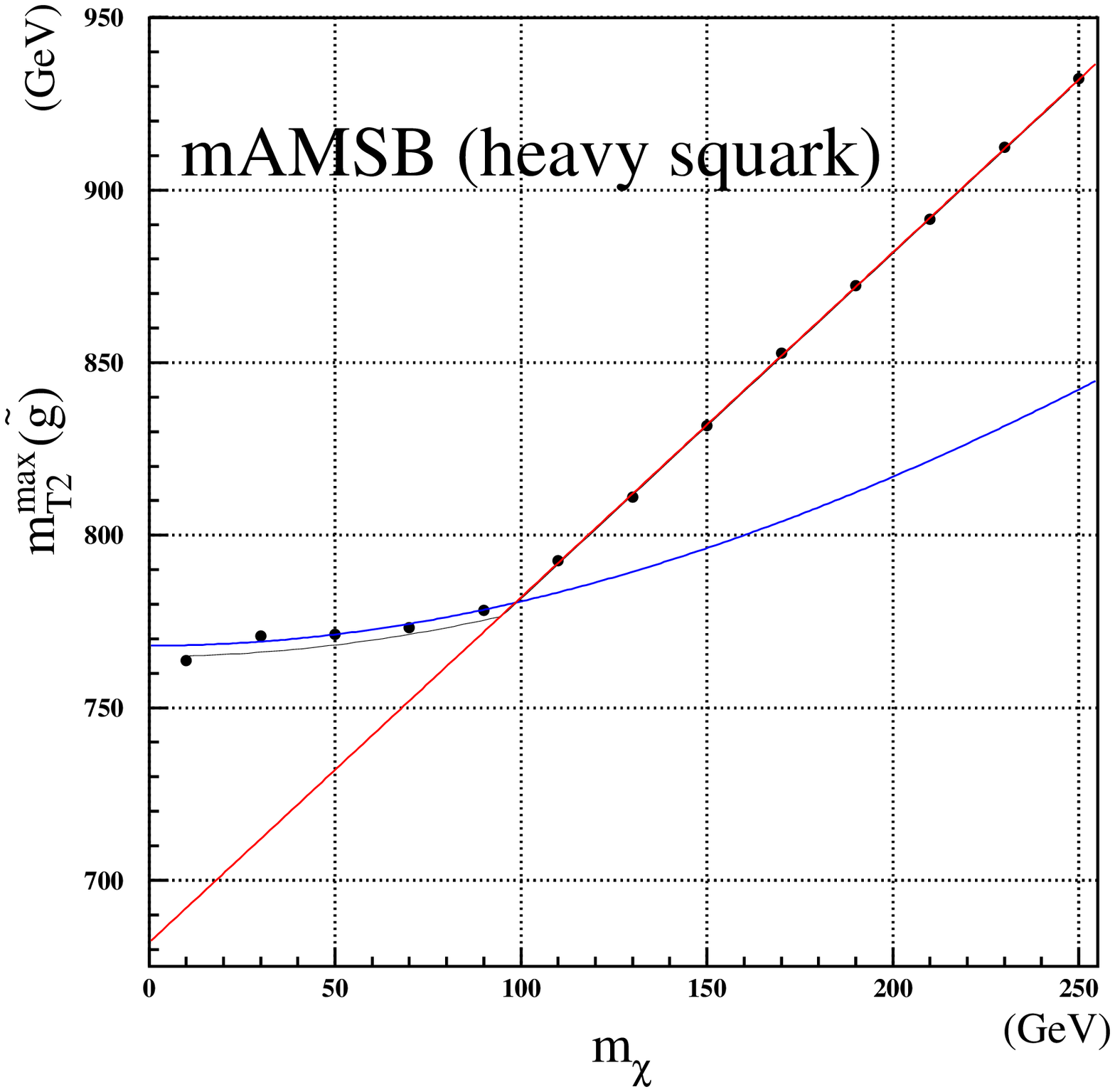, width=6cm, height=6cm}
\end{center}
\caption{(a) Gluino $m_{T2}$ distribution with $m_\chi = 90$ GeV for
AMSB with heavy sfermions, and (b) $m_{T2}^{\rm max}$ as a function
of $m_\chi$ for AMSB with heavy sfermions.} \label{fig:mt2qqx_heavy}
\end{figure}
To obtain a clean signal sample for the gluino $m_{T2}$, we have
imposed the following event selection cuts on the SUSY and SM event
samples.
%
\begin{enumerate}
\item At least 4 jets with $P_{T1,2,3,4} > 200, 150, 100, 50$ GeV.
\item Missing transverse energy $E_T ^{miss} > 250$ GeV.
\item Transverse sphericity $S_T > 0.25$.
\item No b-jets and no leptons.
\end{enumerate}
%
Using the event set passing these selection cuts, we calculate the
gluino $m_{T2}$ for various values of the trial LSP mass $m_\chi$.
Fig. \ref{fig:mt2qqx_heavy} (a) shows the resulting gluino $m_{T2}$
distributions  for the AMSB with $m_\chi=90$ GeV. Fitting the
distribution with a linear function with a linear background, we get
the endpoint value
\begin{eqnarray}
\mbox{AMSB}:\,m_{T2}^{\rm max}(m_\chi=90 )=778.2\pm 2.2 \,\, {\rm
GeV}
\end{eqnarray}
The edge values of $m_{T2}$ obtained in this way
are shown in Fig. \ref{fig:mt2qqx_heavy} (b). Blue and red lines
denote the theoretical curves obtained from (\ref{result1}). Fitting
the data points to these curves, we obtain the following gluino and
LSP masses:
\begin{eqnarray}
\mbox{AMSB}:\, m_{\tilde{g}}=776.5\pm 1.0, \quad
m_{\tilde{\chi}^0_1}=94.9\pm 1.4 \,\, {\rm GeV},
\end{eqnarray}
which are quite close to the true values in (\ref{benchmark}). This
demonstrates that the gluino $m_{T2}$ can be very useful for
measuring the gluino and the LSP masses experimentally in heavier
squark scenario.
%

%
%

Let us now consider the case of lighter squarks,
 $m_{\tilde q}<m_{\tilde g}$,
for which the cascade decay $\tilde g \rightarrow q \tilde q
\rightarrow q q \chi_1^0$ is open.
As an example of superparticle spectra with lighter squarks, we
choose a parameter point (SPS1a \cite{sps1a}) of mSUGRA schemes,
which provides
\begin{eqnarray}
\mbox{mSUGRA with light squarks}: \, m_{\tilde{g}}=613, \quad
m_{\tilde{q}}=525, \quad m_{\tilde{\chi}_1^0}=99 \,\, \mbox{GeV}.
\label{model2}\end{eqnarray} For this mSUGRA point, the production
cross sections for $\tilde g\tilde g$, $\tilde g\tilde q$ and
$\tilde q\tilde q$ pairs are $\sigma(\tilde g\tilde g)\sim 4.2$ pb,
$\sigma(\tilde g\tilde q)\sim 21$ pb, and $\sigma(\tilde q\tilde
q)\sim 9$ pb, respectively. The branching ratio of the signal decay
chain, i.e, $\tilde g\rightarrow \tilde q q\rightarrow
\tilde\chi_1^0 q q$ is $B(\tilde g\rightarrow \tilde\chi_1^0 q q)
\sim 40\%$, while corresponding branching ratios to
$\tilde\chi_2^0$, and $\tilde\chi_1^{\pm}$ are $B(\tilde
g\rightarrow \tilde\chi_2^0 q q)\sim 7\%$, and $B(\tilde
g\rightarrow \tilde\chi_1^\pm q q')\sim 14\%$, respectively. Here,
we assume 30 $fb^{-1}$ of integrated luminosity for the mSUGRA
point.

Similarly to the above AMSB case, we have imposed following event
selection cuts:
%
\begin{enumerate}
\item Missing transverse energy $E_T ^{miss} > 250$ GeV.
\item Transverse sphericity $S_T > 0.25$.
\item No b-jets and no leptons.
\end{enumerate}
\begin{figure}[ht!]
\begin{center}
\epsfig{file=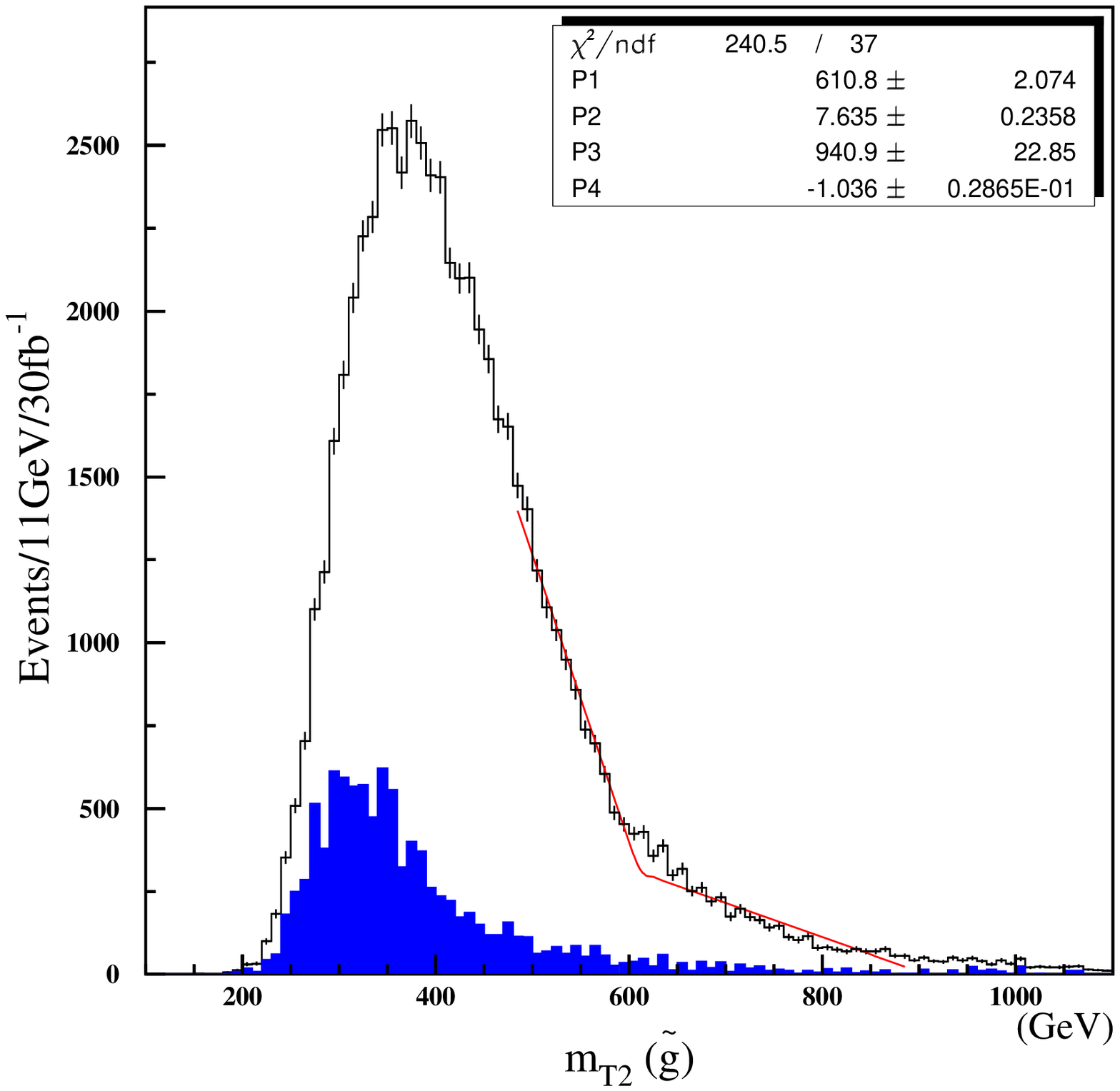, width=6cm, height=6cm}
\epsfig{file=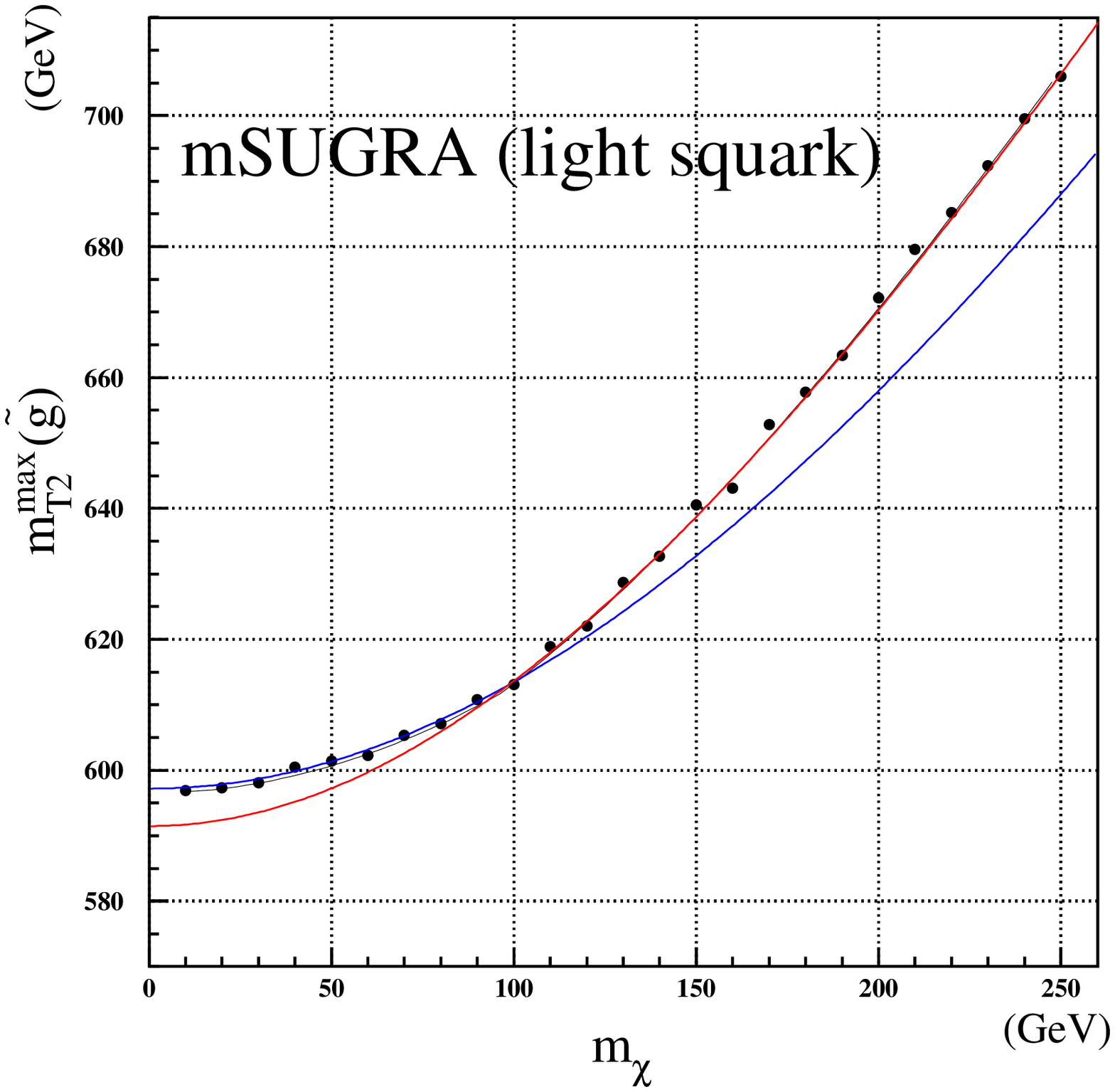, width=6cm, height=6cm}
\end{center}
\caption{Gluino $m_{T2}$ distribution with (a) $m_\chi = 90$ GeV for
the mSUGRA point with light squarks, and (b) $m_{T2}^{\rm max}$ as a
function of $m_\chi$ for mSUGRA with light squarks.}
\label{fig:mt2qqx_light}
\end{figure}
%
%
Again using the Monte Carlo events passing the selection cuts,
we evaluated gluino $m_{T2}$. Fig.\ref{fig:mt2qqx_light} (a) shows
the $m_{T2}$ distribution for trial LSP mass $m_\chi=90$ in mSUGRA
scenario with light squark. Even with the above selection cuts, the
events from the $\tilde g\tilde q$ pair production largely
contribute to the $m_{T2}$ distribution. The contribution from
$\tilde g\tilde q$ pair events provides rather similar shape of
$m_{T2}$ distribution to the one from $\tilde g\tilde g$ events, but
the maximum of $m_{T2}$ from $\tilde g\tilde q$ events is still
smaller than the one from $\tilde g\tilde g$ events.

Fitting the edges of these distributions, we find the endpoint
values:
\begin{eqnarray} \mbox{mSUGRA}:\, m_{T2}^{\rm
max}(m_\chi=90)= 610.8 \pm 2.1 \,\, \mbox{GeV}.
\end{eqnarray}
%
%
Fig.\ref{fig:mt2qqx_light} (b) shows  $m_{T2}^{\rm max}$ as a
function of $m_\chi$. Fitting the data points to the analytic
expression (\ref{result2}), we obtain
\begin{eqnarray}
\mbox{mSUGRA}:\, m_{\tilde{g}}=611.7\pm 2.8, \quad m_{\tilde{q}}=
519.9\pm 2.8, \quad m_{\tilde{\chi}_1^0}=96.3\pm 8.1\,\, {\rm GeV}.
\end{eqnarray}
which are again quite close to the true mass values in
(\ref{model2}). We have performed the same analysis for other
superparticle spectra with $m_{\tilde{q}}< m_{\tilde{g}}$, e.g. a
parameter point of mirage mediation scenario \cite{mirage}, and
found that the gluino mass, squark mass  and LSP mass can be
determined with a similar accuracy.

Here, we emphasize that the above results include only the
statistical uncertainties. There should be various systematic
uncertainties associated with the choice of fit function and the fit
range to determine the endpoint of $m_{T2}$ distribution, which
would affect the result.  Study of such systematic uncertainties,
however, is beyond the scope of this work.

\section{Conclusion}
\label{sec:conclusion}

In this paper, we provided a detailed study of the collider
observable $m_{T2}$ applied for pair-produced superparticles
decaying to visible particles and a pair of invisible LSPs. In
particular, we have derived the analytic expression of the maximum
of $m_{T2}$ over all events ($m_{T2}^{\rm max}$).
 It is noticed that if the decay product of each
superparticle involves more than one visible particle, $m_{T2}^{\rm
max}$ being a function of the {\it trial} LSP mass ${m}_\chi$ has a
kink structure, i.e. a continuous but not differentiable cusp,  at
${m}_\chi=$ true LSP mass, which can be used to determine the mother
superparticle mass and the LSP mass simultaneously. The sharpness of
the kink structure depends on whether the full decay process
involves an intermediate on-shell particle (lighter than the mother
particle) or not. In case without any intermediate on-shell
particle, the kink structure is sharper. In other case with an
intermediate on-shell particle, although the kink structure is
weakened, the $m_{T2}^{\rm max}$-curve can be used to determine the
intermediate particle mass also.

 We also
performed  a Monte-Carlo study of the gluino $m_{T2}$ for some
superparticle spectra in order to examine how well $m_{T2}^{\rm
max}$ can be constructed from collider data. The result of our study
indicates that  the kink structure of $m_{T2}^{\rm max}$ can be
quite useful for the determination of superparticle masses in many
cases, and determine the mother particle mass and the LSP mass quite
accurately in some cases.

\section*{Acknowledgements}

We thank H. D. Kim for useful discussions. This work is supported by
the KRF Grant funded by the Korean Government (KRF-2005-201-C00006),
the KOSEF Grant (KOSEF R01-2005-000-10404-0), and the Center for
High Energy Physics of Kyungpook National University.

\pagebreak

\section*{Appendix A}

In this appendix, we show that $m_{T2}$ of any event in the
laboratory frame is bounded above by another $m_{T2}$ of an event
induced by mother particle pair {\it at rest}. Let us consider
generic event  induced by a symmetric decay of mother particle pair:
\begin{eqnarray} \Phi_i\rightarrow \tilde{\chi}_1^0+ \mbox{visible
particle(s)},
\end{eqnarray}
and let ${\bf p}^{vis(i)}$ ($i=1,2$) denote the total visible
momentum of the decay product of $\Phi_i$ measured in the laboratory
frame.  The corresponding $m_{T2}$ is determined by the visible
transverse momenta ${\bf p}_T^{vis(i)}$ and the visible invariant
masses $m_{vis}^{(i)}$ as defined in (\ref{mt2_def}):
\begin{eqnarray} m_{T2}({\bf p}_T^{vis(i)}, m_{vis}^{(i)},
m_\chi)=\min_{\{{\bf p}_T^{\chi(1)}+{\bf p}_T^{\chi(2)}=-{\bf
p}_T^{vis(1)}-{\bf p}_T^{vis(2)}\}} \big[ \max \big\{ m_T^{(1)},
m_T^{(2)}\big\}\big],
\end{eqnarray}
where \begin{eqnarray} m_T^{(i)}=
\sqrt{m_\chi^2+(m_{vis}^{(i)})^2+2E_T^{vis(i)}E_T^{\chi(i)}-2{\bf
p}_T^{vis(i)}\cdot{\bf p}_T^{\chi (i)}}.
\end{eqnarray}

As the first step, let us perform independent longitudinal boost of
mother particles to make the mother particle pair to move
back-to-back in transverse direction. Note that one can always make
such longitudinal boost for mother particle pair having a vanishing
total transverse momentum in the laboratory frame, and also the
required longitudinal boost of $\Phi_1$ is generically different
from the one of $\Phi_2$.
 Let ${\bf p}^{\prime vis (i)}$ denote the
visible momentum after such longitudinal boost of mother particles.
As ${\bf p}_T$ and $E_T=\sqrt{m^2+|{\bf p}_T|^2}$ are invariant
under longitudinal boost, we obviously have
\begin{eqnarray}
m_{T2}({\bf p}_T^{vis(i)}, m_{vis}^{(i)}, m_\chi) =m_{T2}({\bf
p}^{\prime vis(i)}, m_{vis}^{(i)}, m_\chi). \label{1st}
\end{eqnarray}

To proceed, let us consider the generalized invariant mass $m_{I2}$
defined as follows:
\begin{eqnarray}
m_{I2}({\bf p}^{vis (i)}, m_{vis}^{(i)}, m_\chi)=\min_{\{{\bf
p}^{\chi (1)}+{\bf p}^{\chi (2)}=-{\bf p}^{vis (1)}-{\bf p}^{vis
(2)}\}} \big[\max \big\{m^{(1)}, m^{(2)}\big\}\big],
\end{eqnarray}
where $m^{(i)}$ is the trial invariant mass of the mother particle
$\Phi_i$ obtained for the trial LSP mass $m_\chi$:
\begin{eqnarray} m^{(i)}=
\sqrt{m_\chi^2+(m_{vis}^{(i)})^2+2E^{vis(i)}E^{\chi(i)}-2{\bf
p}^{vis(i)}\cdot{\bf p}^{\chi (i)}},
\end{eqnarray}
and the minimization is performed over the trial LSP momenta
satisfying
\begin{eqnarray}
{\bf p}^{\chi(1)}+{\bf p}^{\chi(2)}=-{\bf p}^{vis(1)}-{\bf
p}^{vis(2)}.
\end{eqnarray}
 As $m_T^{(i)}\leq
m^{(i)}$ for arbitrary visible and trial momenta, one immediately
finds
\begin{eqnarray}
m_{T2}({\bf p}_T^{\prime vis (i)}, m_{vis}^{(i)},m_\chi)\leq
m_{I2}({\bf p}^{\prime vis (i)}, m_{vis}^{(i)},m_\chi), \label{2nd}
\end{eqnarray}
where the equality holds for ${\bf p}^{\prime vis(i)}$ in
$T$-direction. Note that $m_{I2}$ can be considered as an
$(1+3)$-dimensional analogue of the $(1+2)$-dimensional $m_{T2}$.

 The global minimum of
$m^{(i)}$ over the unconstrained ${\bf p}^{\chi (i)}$ is given by
\begin{eqnarray} \big(m^{(i)}\big)_{\rm min}=m_{vis}^{(i)}+m_\chi,
\end{eqnarray}
which occurs when
\begin{eqnarray}
{{\bold p}^{\chi (i)}}=E^{\chi(i)}{\bold p}^{vis (i)}/E^{vis
(i)}=m_\chi{\bold p}^{vis (i)}/m_{vis}^{(i)}.
\end{eqnarray}
Like $m_{T2}$, the generalized invariant mass $m_{I2}$  is also
given by either a balanced solution or an unbalanced solution. If
$m^{(i)}\geq m^{(j)}$  for both $i$ when the trial LSP momenta take
the value giving the unconstrained minimum of $m^{(j)}$ ($j\neq i$),
i.e. if
\begin{eqnarray}
&&m^{(1)}\big|_{{\bf p}^{\chi(1)}=-{\bf p}^{vis(1)}-{\bf
p}^{vis(2)}-\tilde{\bf p}^{\chi(2)}}\,\geq\, m^{(2)}\big|_{{\bf
p}^{\chi(2)}=\tilde{\bf
p}^{\chi(2)}}\,=\,m_{vis}^{(2)}+m_\chi,\nonumber
\\
&&m^{(2)}\big|_{{\bf p}^{\chi(2)}=-{\bf p}^{vis(1)}-{\bf
p}^{vis(2)}-\tilde{\bf p}^{\chi(1)}}\,\geq\, m^{(1)}\big|_{{\bf
p}^{\chi(1)}=\tilde{\bf p}^{\chi(1)}}\,=\,m_{vis}^{(1)}+m_\chi,
\label{bal_mi2_con}
\end{eqnarray}
where
$$\tilde{\bf p}^{\chi(i)}/m_\chi={\bf p}^{vis(i)}/m_{vis}^{(i)},$$ the corresponding  $m_{I2}$ is given by a
balanced solution:
\begin{eqnarray}
m_{I2}=m_{I2}^{\rm bal}= \min_{\{{\bold p}^{\chi(1)}+{\bold
p}^{\chi(2)}=-{\bold p}^{{vis}(1)}-{\bold p}^{{vis}(2)},\,
m^{(1)}=m^{(2)}\}} \,\big[\,m^{(1)}\,\big],
\label{bal_mi2_sol}
\end{eqnarray}
 where the minimization is performed over
${\bf p}^{\chi (i)}$ satisfying
\begin{eqnarray}
m^{(1)}({\bold p}^{{vis}(1)},{\bold p}^{\chi(1)},m_{vis}^{(1)},
m_\chi)&=&m^{(2)}({\bold p}^{{vis}(2)},{\bold
p}^{\chi(2)},m_{vis}^{(2)},
m_\chi),\nonumber \\
{\bold p}^{\chi(1)}+{\bold p}^{\chi(2)}&=&-{\bold
p}^{{vis}(1)}-{\bold p}^{{vis}(2)}. \label{bal_mi2_cons}
\end{eqnarray}

On the other hand, if $m^{(i)} \leq m^{(j)}$ for any $i$ when the
trial LSP momenta take the value giving the unconstrained minimum of
$m^{(j)}$ ($j\neq i$), i.e. if
\begin{eqnarray}
m^{(i)}\big|_{{\bf p}^{\chi(i)}=\tilde{\bf p}^{\chi(i)}}\, \leq\,
m^{(j)}\big|_{{\bf p}^{\chi(j)}=\tilde{\bf p}^{\chi(j)}} \quad
(j\neq i) \label{unbal_mi2_con}
\end{eqnarray}
for the trial LSP momenta given by
\begin{eqnarray} \tilde{\bf p}^{\chi(j)}&=&\tilde{\bf
p}^{\chi(1)}\,=\,m_\chi{\bf p}^{vis(1)}/m_{vis}^{(1)}, \nonumber \\
\tilde{\bf p}^{\chi(2)}&=&-{\bf p}^{vis(1)}-{\bf
p}^{vis(2)}-\tilde{\bf p}^{\chi(1)},
\label{unbal_mi2_rel}\end{eqnarray} or
\begin{eqnarray} \tilde{\bf p}^{\chi(j)}&=&\tilde{\bf
p}^{\chi(2)}\,=\,m_\chi{\bf p}^{vis(2)}/m_{vis}^{(2)}, \nonumber \\
\tilde{\bf p}^{\chi(1)}&=&-{\bf p}^{vis(1)}-{\bf
p}^{vis(2)}-\tilde{\bf p}^{\chi(2)},\end{eqnarray} the corresponding
$m_{I2}$ is given by an unbalanced solution as
\begin{eqnarray} m_{I2}=m^{\rm unbal}_{I2}= m_{vis}^{(j)}+m_\chi
\quad (j=1 \,\, \mbox{or}\,\, 2). \label{unbal_mi2_sol}
\end{eqnarray}

In section 2, we have noticed that $m_{T2}$ is invariant under the
back-to-back boost of ${\bf p}^{vis (i)}$
 along the
direction of the transverse plane $T$, if both ${\bf p}^{vis (1)}$
and ${\bf p}^{vis (2)}$ are in the direction of $T$. It is in fact
straightforward to show that $m_{I2}$ is invariant under
back-to-back boost in general direction for general ${\bf p}^{vis
(i)}$, which corresponds to the $(1+3)$-dimensional version of the
back-to-back boost invariance of $m_{T2}$.
 To show this, let us first note that the invariant masses $m^{(i)}$
and the relations (\ref{bal_mi2_con}) and (\ref{bal_mi2_cons}) are
invariant or covariant under the following back-to-back Lorentz
boost:
\begin{eqnarray}
&&\alpha_1^\mu\rightarrow \Lambda^\mu_\nu(\vec{v})\alpha_1^\nu,\quad
\beta_1^\mu\rightarrow \Lambda^\mu_\nu(\vec{v})\beta_1^\nu,\nonumber \\
&&\alpha_2^\mu\rightarrow \Lambda^\mu_\nu(-\vec{v})\alpha_2^\nu,
\quad \beta_2^\mu\rightarrow \Lambda^\mu_\nu(-\vec{v})\beta_2^\nu,
\label{btb}
\end{eqnarray}
where  $\alpha_i^\mu=(E^{vis (i)}, {\bf p}^{vis(i)})$ are the
visible 4-momenta, $\beta_i^\mu=(E^{\chi(i)},{\bf p}^{\chi (i)})$
are the trial LSP 4-momenta, and
 $\Lambda^\mu_\nu(\vec{v})$ denotes the $(1+3)$-dimensional
Lorentz transformation for generic 3-dimensional boost parameter
$\vec{v}$. This assures that $m_{I2}$ given by a balanced solution
is invariant under generic back-to-back boost, which can be
confirmed by the following explicit form of $m_{I2}^{\rm bal}$:
\begin{eqnarray}
 &&\big(m^{\rm bal}_{I2}\big)^2\,=\, m_\chi^2 + A
 \nonumber \\
 &&\qquad +\, \sqrt{\left(1+ {4
m_\chi^2 \over
2A-\left(m_{vis}^{(1)}\right)^2-\left(m_{vis}^{(2)}\right)^2}\right)
\left(A^2 -\Big(m_{vis}^{(1)}m_{vis}^{(2)}\Big)^2 \right)},
\end{eqnarray}
where $A$ is the Euclidean product of the two visible 4-momenta
$\alpha_1^\mu$ and $\alpha_2^\mu$:\begin{eqnarray} A
=E^{{vis}(1)}E^{{vis}(2)}+{\bold p}^{{vis}(1)}\cdot {\bold
p}^{{vis}(2)}.
\end{eqnarray}
Similarly, the relations (\ref{unbal_mi2_con}) and
(\ref{unbal_mi2_rel}) are covariant under the above back-to-back
Lorentz boost, so $m_{I2}$ given by an unbalanced solution is
invariant also.

In fact, the covariance   of  (\ref{bal_mi2_con}) and
(\ref{unbal_mi2_con}) under the back-to-back boost (\ref{btb}),
which we have used to show the invariance of $m_{I2}$, is not so
obvious. The easiest way to see their covariance is to consider the
boundary between (\ref{bal_mi2_con}) and (\ref{unbal_mi2_con}),
which corresponds to the visible momenta satisfying
\begin{eqnarray}
m^{(1)}({\bold p}^{{vis}(1)},{\bold p}^{\chi(1)},m_{vis}^{(1)},
m_\chi)=m^{(2)}({\bold p}^{{vis}(2)},{\bold
p}^{\chi(2)},m_{vis}^{(2)}, m_\chi), \label{bound1}
\end{eqnarray}
for the trial momenta given by
\begin{eqnarray}
{\bf p}^{\chi(1)}&=&m_\chi{\bf p}^{vis(1)}/m_{vis}^{(1)}, \nonumber \\
{\bold p}^{\chi(2)}&=&-{\bold p}^{\chi(1)}-{\bold
p}^{{vis}(1)}-{\bold p}^{{vis}(2)}, \label{bound2}
\end{eqnarray}
or by
\begin{eqnarray}
{\bf p}^{\chi(2)}&=&m_\chi{\bf p}^{vis(2)}/m_{vis}^{(2)},
\nonumber \\
{\bold p}^{\chi(1)}&=&-{\bold p}^{\chi(2)}-{\bold
p}^{{vis}(1)}-{\bold p}^{{vis}(2)}.\label{bound3}\end{eqnarray} If
some momenta satisfy (\ref{bound1}) and (\ref{bound2}), or
(\ref{bound1}) and (\ref{bound3}), their back-to-back boost satisfy
also the same relations. This means that the boundary is invariant
under the back-to-back boost, thus there can not be any crossing of
the boundary caused by the back-to-back boost. Therefore, if some
visible momenta satisfy (\ref{bal_mi2_con}) or
(\ref{unbal_mi2_con}), their back-to-back boost satisfy the same
condition, so (\ref{bal_mi2_con}) and (\ref{unbal_mi2_con}) are
covariant under the back-to-back boost (\ref{btb}). The same
argument can be used to show that (\ref{bal_region}) and
(\ref{con_unbalanced}) are covariant under the $(1+2)$-dimensional
transformation (\ref{tr_boost}).

In the above, we have argued that $m_{I2}$ is invariant under
generic back-to-back Lorentz boost, which
leads to
\begin{eqnarray}
m_{I2}({\bf p}^{\prime vis(i)}, m_{vis}^{(i)}, m_\chi) =m_{I2}({\bf
q}^{vis(i)},m_{vis}^{(i)},m_\chi) \label{3rd}
\end{eqnarray} for ${\bf
q}^{vis(i)}$ obtained by  arbitrary back-to-back boost of ${\bf
p}^{\prime vis(i)}$. By definition,  ${\bf p}^{\prime vis(i)}$ is
the $i$-th visible momentum  after
 the independent longitudinal boosts making
the mother particle pair to move back-to-back in the  direction of
$T$. One can then choose an appropriate back-to-back boost for which
${\bf q}^{vis(i)}$ corresponds to the $i$-th visible momentum
measured in the rest frame of its mother particle.

Let $T^\prime$ denote the transverse plane spanned by  ${\bf q}^{vis
(1)}$ and ${\bf q}^{vis (2)}$, and $z^\prime$ denote its normal
direction. Then, by definition ${\bf q}^{vis (i)}_{z^\prime}=0$, and
it is straightforward to minimize $\max \big\{ m^{(1)}, m^{(2)}
\big\}$ over the $z^\prime$-component of the trial LSP momentum:
\begin{eqnarray} &&\min_{\{{\bf p}^{\chi (1)}+{\bf p}^{\chi
(2)}=-{\bf q}^{vis (1)}-{\bf q}^{vis
(2)}\}} \big[ \max \big\{ m^{(1)}, m^{(2)} \big\} \big] \nonumber\\
&=& \min_{\{{\bf p}_{T^\prime}^{\chi (1)}+{\bf p}_{T^\prime}^{\chi
(2)}=-{\bf q}_{T^\prime}^{vis (1)}-{\bf q}_{T^\prime}^{vis (2)}\}}
\big[ \max \big\{m^{(1)}, m^{(2)} \big\}_{{\bf p}^{\chi
(1)}_{z^\prime}={\bf p}^{\chi (2)}_{z^\prime}=0}
\big] \nonumber \\
&=& \min_{\{{\bf p}_{T^\prime}^{\chi (1)}+{\bf p}_{T^\prime}^{\chi
(2)}=-{\bf q}_{T^\prime}^{vis (1)}-{\bf q}_{T^\prime}^{vis (2)}\}}
\big[ \max \big\{ m_{T^\prime}^{(1)}, m_{T^\prime}^{(2)}
\big\}\big],
\end{eqnarray}
and thus
\begin{eqnarray}
m_{I2}({\bf q}^{vis(i)},m_{vis}^{(i)},m_\chi) = m_{T^\prime 2}({\bf
q}_{T^\prime}^{vis(i)},m_{vis}^{(i)},m_\chi ). \label{4th}
\end{eqnarray}
Combining (\ref{1st}), (\ref{2nd}), (\ref{3rd}) and (\ref{4th}), we
finally obtain
\begin{eqnarray}
m_{T2}({\bf p}^{vis(i)}, m_{vis}^{(i)}, m_\chi) \leq m_{T^\prime
2}({\bf q}^{vis(i)}, m_{vis}^{(i)}, m_\chi)
\end{eqnarray}
for arbitrary symmetric decay of mother particle pair having a
vanishing total transverse momentum in the direction of $T$, where
the equality holds when $T^\prime=T$. Therefore, $m_{T2}$ of any
event induced by mother particle pair having a vanishing total
transverse momentum in the laboratory frame is {\it bounded  above}
by another $m_{T2}$ of an event induced by mother particle pair {\it
at rest}.

\section*{Appendix B.}

In this appendix, we discuss in detail the $m_{T2}$ of the events
associated with  the decay of mother particle pair {\it at rest}
with visible momenta in transverse direction. Such $m_{T2}$
corresponds to $m_{T^\prime 2}({\bf
q}^{vis(i)},m_{vis}^{(i)},m_\chi)$, where ${\bf q}^{vis(i)}$ denotes
the $i$-th visible momenta measured in the rest frame of its mother
particle, and $T^\prime$ is the transverse plane spanned by ${\bf
q}^{vis(1)}$ and ${\bf q}^{vis(2)}$. Since $|{\bf q}^{vis(i)}|$ is
determined as
\begin{eqnarray}
|{\bf q}^{vis(i)}|&=&\frac{1}{2
\tilde{m}}\left[\Big((\tilde{m}+m_{vis}^{(i)})^2-m_{\tilde\chi_1^0}^2\Big)
\Big((\tilde{m}-m_{vis}^{(i)})^2-m_{\tilde\chi_1^0}^2\Big)\right]^{1/2},
\label{q_mag}
\end{eqnarray}
where $\tilde{m}$ is the mother particle mass and
$m_{\tilde{\chi}_1^0}$ is the LSP mass,
 $m_{T^\prime 2}$ appears as
a function of  the three event variables $m_{vis}^{(1)},
m_{vis}^{(2)}, \theta=$ the angle between ${\bf q}^{vis(1)}$ and
${\bf q}^{vis(2)}$, and also the trial LSP mass $m_\chi$:
\begin{eqnarray}
m_{T^{\prime}2}({\bf
q}^{vis(i)},m_{vis}^{(i)},m_\chi)\,\equiv\,{\cal F}(m_{vis}^{(i)},
\theta, m_\chi).
\end{eqnarray}
In section 2, we discussed the behavior of ${\cal F}$ over the
one-dimensional event space with
$m_{vis}^{(1)}=m_{vis}^{(2)}=m_{vis}$ and $\theta=0$.
Here we generalize the analysis to the full 3-dimensional event
space spanned by  $\{m_{vis}^{(i)},\theta\}$, and show
\begin{eqnarray}
&&\frac{\partial {\cal F}}{\partial\theta}\leq 0 \quad\,\,\mbox{for
any}\,\, m_{vis}^{(i)},m_\chi,\,\,\mbox{and}\,\,0\leq\theta\leq\pi, \nonumber \\
&&\left.\frac{\partial {\cal F}}{\partial
m_{vis}^{(i)}}\right|_{\theta=0}=\left\{\begin{array}{ll} \,\geq 0
\quad \mbox{for} \,\, m_{\chi}>m_{\tilde{\chi}_1^0} \,\,\mbox{and any}\,\, m_{vis}^{(i)} \\
\,\leq 0\quad \mbox{for} \,\,
m_{\chi}<m_{\tilde{\chi}_1^0}\,\,\mbox{and any}\,\, m_{vis}^{(i)},
\end{array}\right. \end{eqnarray}
and thus the global maximum of ${\cal F}$ over all
$\{m_{vis}^{(i)},\theta\}$ is given by
\begin{eqnarray}
{\cal F}^{\rm max}(m_{\chi})=\left\{\begin{array}{ll} {\cal F}^{\rm
max}_<
  \quad \mbox{for} \,\,
m_{\chi}<m_{\tilde{\chi}_1^0} \\ {\cal F}^{\rm max}_{>}\quad
\mbox{for} \,\, m_{\chi}>m_{\tilde{\chi}_1^0},
\end{array}\right. \end{eqnarray}
where \begin{eqnarray}
 {\cal F}^{\rm max}_< &=&{\cal
F}(m_{vis}^{(1)}=m_{vis}^{\rm min}, m_{vis}^{(2)}=m_{vis}^{\rm
min},\theta=0,m_\chi),\nonumber \\{\cal F}^{\rm max}_{>} &=&{\cal
F}(m_{vis}^{(1)}=m_{vis}^{\rm max}, m_{vis}^{(2)}=m_{vis}^{\rm
max},\theta=0,m_\chi).
\end{eqnarray}

As discussed in section 2,  $m_{T2}$ is given by either a balanced
solution or unbalanced solution, depending upon whether the
condition (\ref{bal_region}) is satisfied or not:
\begin{eqnarray}
{\cal F}=\left\{\begin{array}{ll} {\cal F}^{\rm bal}
\quad&\mbox{for} \,\, {\bf q}^{vis(i)}\,\,\mbox{in the balanced
domain},\nonumber
\\
{\cal F}^{\rm unbal} \quad &\mbox{otherwise},
\end{array}
\right.\end{eqnarray} where
\begin{eqnarray}
\label{app-bal}
\big({\cal F}^{\rm bal}\big)^2&=& m_\chi^2 + A \\
&+& \sqrt{\left(1+ {4 m_\chi^2 \over 2
A-(m_{vis}^{(1)})^2-(m_{vis}^{(2)})^2 }\right) \left(A^2
-(m_{vis}^{(1)})^2 (m_{vis}^{(2)})^2 \right)}, \label{f_bal}
\\
{\cal F}^{\rm unbal}&=& m_\chi+m_{vis}^{(i)} \quad (i=1
\,\,\mbox{or}\,\, 2)
\end{eqnarray}
for
\begin{eqnarray}
A =E^{vis(1)} E^{vis(2)} + |{\bold q}^{vis(1)}||{\bold q}^{vis(2)}|
{\rm cos}\theta,
\end{eqnarray}
with $|{\bold q}^{vis(i)}|$  given by (\ref{q_mag}) and
$E^{vis(i)}=\sqrt{|{\bf q}^{vis(i)}|^2+(m_{vis}^{(i)})^2}$. Here the
balanced domain corresponds to  the event set satisfying
(\ref{bal_region}), and  ${\cal F}^{\rm bal}$ is obtained from
(\ref{balanced}) with ${\bf p}_T^{vis(i)}={\bf q}^{vis(i)}$. Note
that ${\bf q}^{vis(i)}={\bf q}_{T^\prime}^{vis(i)}$ and
$A=A_{T^\prime}$ according to the definition of $T^\prime$.



\subsection*{B.1 \, ${\cal F}$ vs. $\theta$}

Let us show that ${\cal F}$ has its maximum at $\theta=0$ for given
values of $m_{vis}^{(i)}$ and $m_\chi$. If
$m_{vis}^{(1)}=m_{vis}^{(2)}\equiv m_{vis}$, one easily finds ${\cal
F}$ is given by a balanced solution of the form
\begin{eqnarray}
{\cal F}^{\rm
bal}=\sqrt{\frac{A+m_{vis}^2}{2}}+\sqrt{\frac{A-m_{vis}^2+2m_\chi^2}{2}}
\end{eqnarray}
which obviously has its maximum at $\theta=0$ for given values of
$m_{vis}^{(i)}$ and $m_\chi$.

To examine the case with $m_{vis}^{(1)} \neq m_{vis}^{(2)}$, let us
consider
\begin{eqnarray}
\frac{\partial ({\cal F}^{\rm
bal})^2}{\partial\theta}=-\frac{|{\bold q}^{vis(1)}| |{\bold
q}^{vis(2)}|\sin\theta}{
\sqrt{B^3C}\chi}\left[(AB-C)\chi^2+\sqrt{B^{3}C}\chi+BC\right],
\label{deriv1}
\end{eqnarray}
where
\begin{align*}
A &\equiv E^{vis(1)} E^{vis(2)} + |{\bold q}^{vis(1)}||{\bold
q}^{vis(2)}| \cos\theta,\\
B & \equiv 2A-(m_{vis}^{(1)})^2-(m_{vis}^{(2)})^2,\\
C & \equiv A^{2}-(m_{vis}^{(1)}m_{vis}^{(2)})^2, \\
\chi &\equiv \sqrt{B+4m_\chi^2}.
\end{align*}
It is  straightforward to find that $A>0$, $B>0$ and $C>0$ when
$m_{vis}^{(1)}\neq m_{vis}^{(2)}$. Then ${\cal F}^{\rm bal}$ has an
extremum at $\theta=0,\pi$ and also at $\theta=\theta_0$ for which
\begin{eqnarray}
f &\equiv& (AB-C)\chi^2+\sqrt{B^{3}C}~\chi+BC \nonumber \\
&=&
\big(\,(A-(m_{vis}^{(1)})^2)\chi+\sqrt{BC}\,\big)\big(\,(A-(m_{vis}^{(2)})^2)\chi+\sqrt{BC}\,\big)
=0,
\end{eqnarray}
where we have used
$AB-C=(A-(m_{vis}^{(1)})^2)(A-(m_{vis}^{(2)})^2)$. Here we will
discuss only the case with $m_{vis}^{(1)} > m_{vis}^{(2)}$ since the
result for $m_{vis}^{(1)} < m_{vis}^{(2)}$ can be obtained by
interchanging $m_{vis}^{(1)}$ and $m_{vis}^{(2)}$.

If $AB-C\geq 0$, $f>0$ for $\chi\geq 0$, and thus
\begin{eqnarray}
\frac{\partial {\cal F}^{\rm bal}}{\partial\theta}\leq 0 \quad
\mbox{for any}\,\, m_{vis}^{(i)}, \theta, m_\chi \,\,
\mbox{with}\,\, AB-C\geq 0.
\end{eqnarray}
 As
${\cal F}^{\rm unbal}=m_\chi+m_{vis}^{(1)}$ is independent of
$\theta$, this means that $\partial{\cal F}/\partial\theta\leq 0$,
thus ${\cal F}$ has its maximum at $\theta=0$ for
$m_{vis}^{(i)},m_\chi$ giving $AB-C\geq 0$.

In other case with  $AB-C<0$, one finds $f\geq 0$ for $0<
\chi\leq\chi_0$, while $f< 0$ for $\chi>\chi_0$, where
\begin{eqnarray}
\chi_0=\frac{\sqrt{BC}}{(m_{vis}^{(1)})^2-A}. \label{fzero}
\end{eqnarray}
Since $\theta_0$ corresponds to the value of $\theta$ for which
$\chi=\chi_0$ (for given values of $m_{vis}^{(i)}$ and $m_\chi$),
this implies
 \begin{eqnarray} f= \left\{ \begin{array} {ll}
\,\geq 0 \quad \mbox{for}\,\, 0< \theta\leq \theta_0, \\
\, <0\quad  \mbox{for}\,\, \theta_0< \theta <\pi, \end{array}
\right.
\end{eqnarray}
and thus  ${\cal F}^{\rm bal}$ decreases as $\theta$ varies from
zero to $\theta_0$, while it increases as $\theta$ varies from
$\theta_0$ to $\pi$. On the other hand, $\chi\equiv
\sqrt{B+4m_\chi^2}=\chi_0$ leads to
\begin{eqnarray}
A\big|_{\theta=\theta_0}=\frac{m_{vis}^{(1)}\big(2m_{vis}^{(1)}m_\chi+(m_{vis}^{(1)})^2+
(m_{vis}^{(2)})^2)\big)}{2(m_\chi+m_{vis}^{(1)})}
\end{eqnarray}
for which
\begin{eqnarray}
{\cal F}^{\rm bal}\big|_{\theta=\theta_0} =m_\chi+m_{vis}^{(1)}=
{\cal F}^{\rm unbal}.
\end{eqnarray}
This means that $\theta=\theta_0$ corresponds to the boundary
between the balanced domain and the unbalanced domain. One can show
also ${\cal F}^{\rm bal}\geq  {\cal F}^{\rm unbal}$ for $\theta\geq
\theta_0$, where the equality holds for $\theta=\theta_0$, implying
that ${\cal F}$ is given by ${\cal F}^{\rm unbal}$ for
 $\theta_0<\theta\leq\pi$, while it is given by
 ${\cal F}^{\rm bal}$ for $0\leq \theta\leq \theta_0$.
Again, combined with that ${\cal F}^{\rm unbal}$ is independent of
$\theta$,  these observations lead to \begin{eqnarray}
\frac{\partial{\cal F}}{\partial\theta}\leq 0 \quad\mbox{for} \,\,
AB-C<0,
\end{eqnarray}
 thus ${\cal F}$ has
its maximum at $\theta=0$ for  $m_{vis}^{(i)},m_\chi$ giving
$AB-C<0$ also.

\subsection*{B.2 ${\cal F}$ vs. $m_{vis}^{(i)}$:}

Let us now examine the dependence of ${\cal F}$ on $m_{vis}^{(i)}$.
 As we are interested in the
events giving ${\cal F}^{\rm max}$,  we will fix $\theta=0$ in the
following. We then have
\begin{eqnarray}
\frac{\partial ({\cal F}^{\rm bal})^2}{\partial (m_{vis}^{(1)})^2}
=\frac{g}{2\sqrt{B^{3}C}\chi},
\end{eqnarray}
where
\begin{eqnarray}
g \equiv (BC^{\prime}-B^{\prime}C)\chi^{2}+\sqrt{B^{3}C}(B^{\prime}
+1)\chi+B^{\prime}BC
\end{eqnarray}
with
\begin{eqnarray}
A^{\prime} \equiv \frac{\partial A}{\partial (m_{vis}^{(1)})^2},~~
B^{\prime} \equiv \frac{\partial B}{\partial (m_{vis}^{(1)})^2},~~
C^{\prime} \equiv \frac{\partial C}{\partial (m_{vis}^{(1)})^2}.
\end{eqnarray}

The equation $g=0$ can have the following solutions:
\begin{eqnarray}
\label{solmp}
\chi_{1}&=&
\frac{\sqrt{BC}(1-2A^{\prime})}{2A^{\prime}(A-(m_{vis}^{(1)})^2)+A-(m_{vis}^{(2)})^2},
\nonumber \\
\chi_{2}&=& \frac{\sqrt{BC}}{(m_{vis}^{(2)})^2-A},
\end{eqnarray}
where we have used
\begin{eqnarray}
BC^{\prime}-B^{\prime}C=(A-(m_{vis}^{(2)})^2)\left(2A^{\prime}
(A-(m_{vis}^{(1)})^2)+A-(m_{vis}^{(2)})^2\right) \nonumber
\end{eqnarray}
and
\begin{eqnarray}
\left((m_{vis}^{(1)})^2-(m_{vis}^{(2)})^2\right)A^{\prime}-A+(m_{vis}^{(2)})^2\,<\,0
\quad\mbox{for}\,\,m_{vis}^{(1)} \neq m_{vis}^{(2)}. \nonumber
\end{eqnarray}
If $m_{vis}^{(1)} \neq m_{vis}^{(2)}$, we have also
\begin{eqnarray}
 B^{\prime}=(2A^{\prime}-1)\,<\,0,\nonumber \\
2A^{\prime}(A-(m_{vis}^{(1)})^2)+A-(m_{vis}^{(2)})^2 \,>\, 0,
\label{useful}
\end{eqnarray}
and thus $\chi_{1}$ is always positive, while the sign of $\chi_{2}$
is determined by the sign of $A-(m_{vis}^{(2)})^2$.

If $m_{vis}^{(1)}>m_{vis}^{(2)}$, we have ${\cal F}^{\rm
unbal}=m_\chi+m_{vis}^{(1)}$, and thus
\begin{eqnarray}
{\frac{\partial {\cal F}^{\rm unbal}}{\partial m_{vis}^{(1)}}}=1
\quad \mbox{for} \,\,m_{vis}^{(1)}>m_{vis}^{(2)}.
\label{resulta}\end{eqnarray} As for the behavior of ${\cal F}^{\rm
bal}$, since $A-(m_{vis}^{(2)})^2>0$ for
$m_{vis}^{(1)}>m_{vis}^{(2)}$,  only $\chi=\chi_{1}$ can be a
physical solution of $g=0$, for which
\begin{eqnarray}
\sqrt{B+4m_\chi^{2}}=\frac{\sqrt{BC}(1-2A^{\prime})}{2A^{\prime}(A-(m_{vis}^{(1)})^2)
+A-(m_{vis}^{(2)})^2}. \label{mxmc}
\end{eqnarray}
This equation is solved by $m_\chi=m_{\tilde\chi_1^0}$, which means
$g=0$ at  $m_\chi=m_{\tilde\chi_1^0}$. It is also straightforward to
find that  $g>0$ for $m_\chi>m_{\tilde\chi_1^0}$ and $g<0$ for
$m_\chi<m_{\tilde\chi_1^0}$, leading to
\begin{eqnarray}
{\frac{\partial {\cal F}^{\rm bal}}{\partial
m_{vis}^{(1)}}}=\left\{\begin{array}{ll} \,\leq\,0 \quad
&\mbox{for}\,\, m_{\chi}<m_{\tilde{\chi}_{1}^{0}}\,\,\mbox{and}
\,\,\,m_{vis}^{(1)}>m_{vis}^{(2)},
\\
\,\geq\,0 \quad &\mbox{for}\,\,
m_{\chi}>m_{\tilde{\chi}_{1}^{0}}\,\,\mbox{and}
\,\,\,m_{vis}^{(1)}>m_{vis}^{(2)}.
\end{array}\right.
\label{resultb}
\end{eqnarray}
On the other hand, if $m_\chi<m_{\tilde\chi_1^0}$ and $\theta=0$,
${\cal F}$ is always given by ${\cal F}^{\rm bal}$.  We then find
from (\ref{resulta}) and (\ref{resultb}) that
\begin{eqnarray}
{\frac{\partial {\cal F}}{\partial
m_{vis}^{(1)}}}=\left\{\begin{array}{ll} \,\leq\,0 \quad
&\mbox{for}\,\, m_{\chi}<m_{\tilde{\chi}_{1}^{0}}\,\,\mbox{and}
\,\,\,m_{vis}^{(1)}>m_{vis}^{(2)},
\\
\,\geq\,0 \quad &\mbox{for}\,\,
m_{\chi}>m_{\tilde{\chi}_{1}^{0}}\,\,\mbox{and}
\,\,\,m_{vis}^{(1)}>m_{vis}^{(2)}.
\end{array}\right.
\end{eqnarray}

In other case with $m_{vis}^{(1)}<m_{vis}^{(2)}$,
 we have ${\cal F}^{\rm
unbal}=m_\chi+m_{vis}^{(2)}$, so
\begin{eqnarray}
{\frac{\partial {\cal F}^{\rm unbal}}{\partial m_{vis}^{(1)}}}=0
\quad \mbox{for} \,\,m_{vis}^{(1)}<m_{vis}^{(2)}.
\label{resultc}\end{eqnarray} In this case, $A-(m_{vis}^{(2)})^2$
can be either positive or negative.
If it is positive, again $\chi=\chi_1$ is the only solution for
$g=0$. Then, one can repeat the analysis leading to (\ref{resultb}),
and find
\begin{eqnarray}
{\frac{\partial {\cal F}}{\partial
m_{vis}^{(1)}}}=\left\{\begin{array}{ll} \,\leq\,0 \quad
&\mbox{for}\,\, m_{\chi}<m_{\tilde{\chi}_{1}^{0}},\,\,
m_{vis}^{(1)}<m_{vis}^{(2)}, \,\, A-(m_{vis}^{(2)})^2>0
\\
\,\geq\,0 \quad &\mbox{for}\,\, m_{\chi}>m_{\tilde{\chi}_{1}^{0}},
\,\,m_{vis}^{(1)}<m_{vis}^{(2)},\,\,A-(m_{vis}^{(2)})^2>0
\end{array}\right.
\label{resultd}
\end{eqnarray}

\begin{figure}[ht!]
\begin{center}
\epsfig{figure=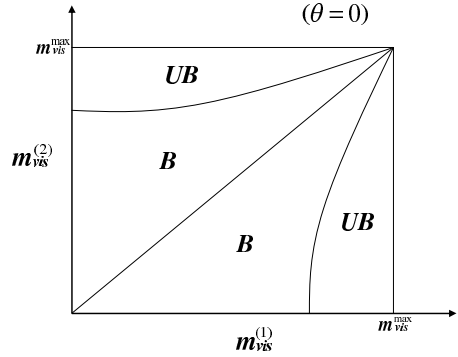,width=7cm,height=7cm}
\end{center}
\caption{Division of the ($m_{vis}^{(1)},m_{vis}^{(2)}$) into the
balanced solution region (B) and the unbalanced region (UB). }
\label{fig:m1m2}
\end{figure}

However, if $A-(m_{vis}^{(2)})^2<0$, both $\chi=\chi_1$ and
$\chi=\chi_2$  become a good solution of $g=0$, and $\chi_2>\chi_1$.
Again, $\chi=\chi_1$ is obtained if and only if
$m_\chi=m_{\tilde\chi_1^0}$. As for the other solution
$\chi=\chi_2$, i.e.
\begin{eqnarray}
 \sqrt{B+4m_\chi^2}=\frac{\sqrt{BC}}{(m_{vis}^{(2)})^2-A},
\label{gzero22}
\end{eqnarray}
it is straightforward to find that it gives rise to
\begin{eqnarray}
{\cal F}^{\rm bal}\big|_{\chi=\chi_2}=m_\chi+m_{vis}^{(2)}={\cal
F}^{\rm unbal},
\end{eqnarray}
which means that $\chi=\chi_2$ corresponds to the boundary between
the balanced domain  and the unbalanced domain.
 In Fig.
\ref{fig:m1m2}, we depict the 2-dimensional event space of
$(m_{vis}^{(1)},m_{vis}^{(2)})$ for $\theta=0$ and $m_{\chi}>
m_{\tilde{\chi}^0_1}$, showing the balanced domain ($B$) and the
unbalanced domain ($UB$).

The above observation implies that ${\cal F}$ is always given by
${\cal F}^{\rm bal}$ if $\chi<\chi_1$, for which
$m_\chi<m_{\tilde\chi_1^0}$.
 One can show that $g<0$ in such case, so
\begin{eqnarray}
{\frac{\partial {\cal F}}{\partial m_{vis}^{(1)}}} \leq 0 \quad
\mbox{for}\,\,m_{\chi}<m_{\tilde{\chi}_{1}^{0}},
\,\,m_{vis}^{(1)}<m_{vis}^{(2)},\,\, A-(m_{vis}^{(2)})^2<0.
\label{resulte}
\end{eqnarray}
However, if $\chi\geq \chi_1$, for which $m_\chi\geq
m_{\tilde{\chi}_1^0}$, the function $g$ can have either sign.
 We find that $g\geq 0$ for $\chi_1\leq \chi\leq \chi_2$, while
$g\leq 0$ for $\chi\geq \chi_2$. As $\chi=\chi_2$ corresponds to the
boundary between the balanced domain and the unbalanced domain, this
means that $g\geq 0$ when ${\cal F}$ is given by ${\cal F}^{\rm
bal}$, so
\begin{eqnarray}
&&\qquad\qquad\qquad {\frac{\partial {\cal F}}{\partial
m_{vis}^{(1)}}}= {\frac{\partial
{\cal F}^{\rm bal}}{\partial m_{vis}^{(1)}}}\geq 0 \nonumber \\
&& \mbox{for}\,\, m_{\chi}>m_{\tilde{\chi}_{1}^{0}},\,\,
m_{vis}^{(1)}<m_{vis}^{(2)}, \,\, A-(m_{vis}^{(2)})^2<0, \,\,
\chi_1\leq\chi\leq \chi_2. \label{resultf}
\end{eqnarray}
If $\chi\geq \chi_2$ with $m_{vis}^{(1)}<m_{vis}^{(2)}$, ${\cal F}$
is given by ${\cal F}^{\rm unbal}=m_\chi+m_{vis}^{(2)}$, so
\begin{eqnarray}
&&\qquad\qquad\qquad{\frac{\partial {\cal F}}{\partial
m_{vis}^{(1)}}}= {\frac{\partial {\cal F}^{\rm unbal}}{\partial
m_{vis}^{(1)}}}=0
\nonumber \\
&& \mbox{for}\,\, m_{\chi}>m_{\tilde{\chi}_{1}^{0}},\,\,
m_{vis}^{(1)}<m_{vis}^{(2)}, \,\, A-(m_{vis}^{(2)})^2<0, \,\,
\chi\geq \chi_2. \label{resultg}
\end{eqnarray}

\begin{figure}[ht!]
\begin{center}
\epsfig{figure=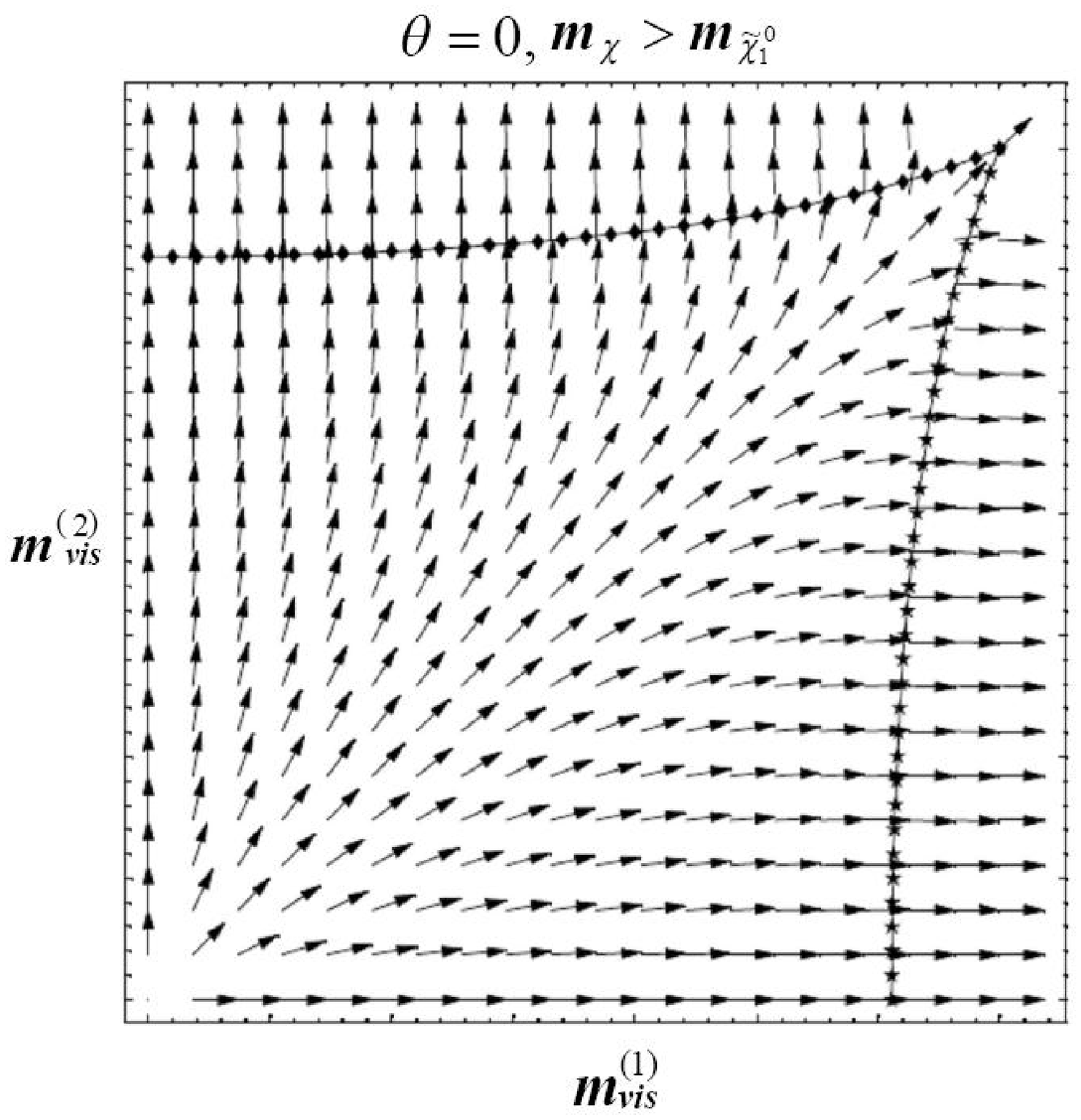,width=7cm,height=7cm}
\epsfig{figure=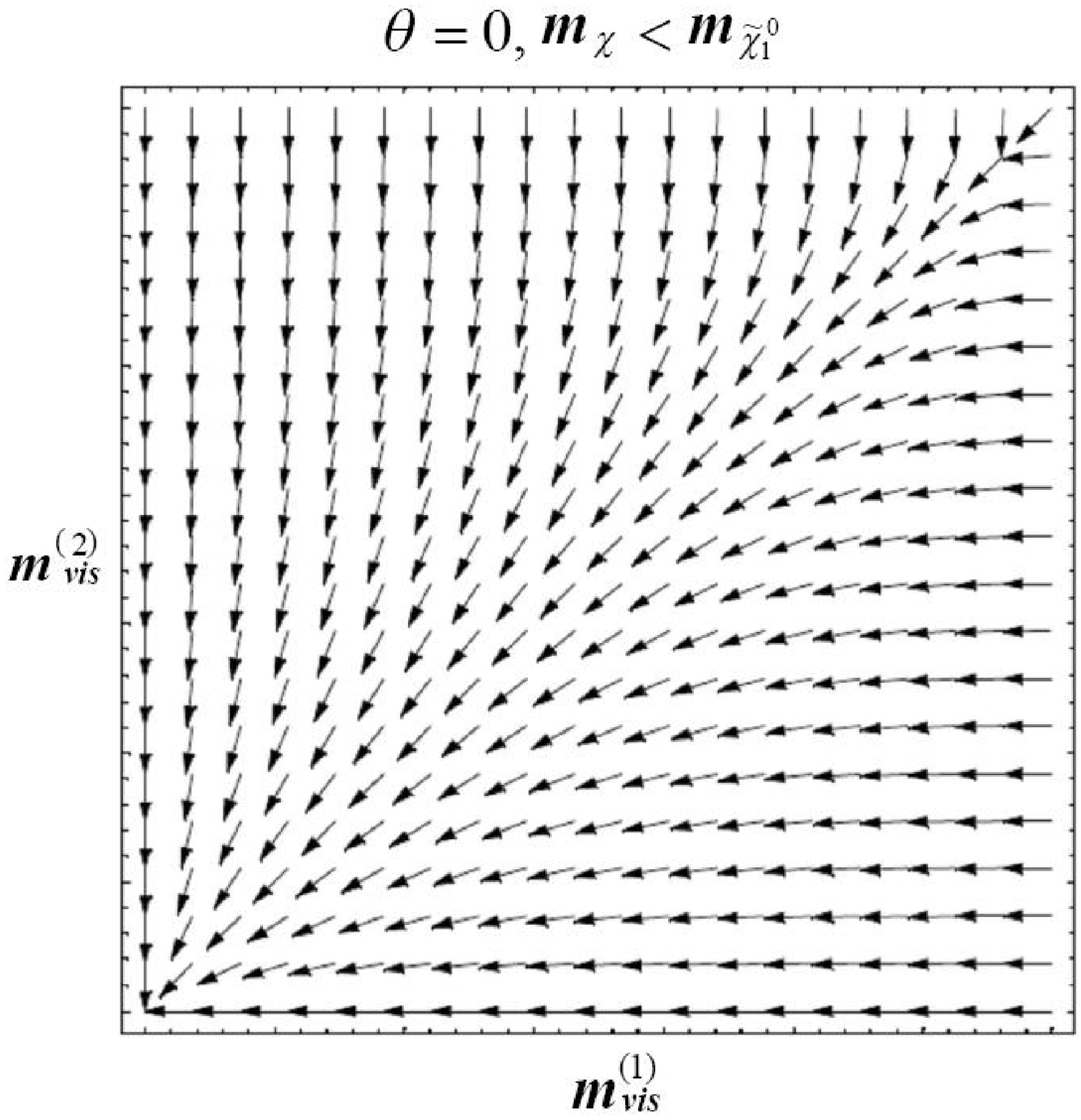,width=7cm,height=7cm}
\end{center}
\caption{The gradient  of ${\cal F}$  for (a) $m_\chi
> m_{\tilde{\chi}_1^0}$ and $\theta=0$, (b) $m_\chi
<m_{\tilde{\chi}_1^0}$ and $\theta=0$. }\label{fig:gradients}
\end{figure}

Combining all of the above results together, and also taking into
account that ${\cal F}$ is invariant under the exchange of
$m_{vis}^{(1)}$ and $m_{vis}^{(2)}$, we finally obtain
\begin{eqnarray}
\left.{\frac{\partial {\cal F}}{\partial
m_{vis}^{(i)}}}\right|_{\theta=0} =\left\{ \begin{array} {ll} \leq 0
\quad \mbox{for}\,\,m_{\chi}<m_{\tilde{\chi}_{1}^{0}}\,\,\mbox{and any}\,\, m_{vis}^{(i)}\\
\geq 0\quad\mbox{for}\,\,
m_{\chi}>m_{\tilde{\chi}_{1}^{0}}\,\,\mbox{and any}\,\,
m_{vis}^{(i)}.
\end{array}\right.
\end{eqnarray} In Fig.(\ref{fig:gradients}), we depict the pattern of the 2-d vector field
$\partial{\cal F}/\partial m_{vis}^{(i)}$ for both cases with
$m_\chi<m_{\tilde{\chi}_1^0}$ ($\chi<\chi_1$) and
$m_\chi>m_{\tilde{\chi}_1^0}$ ($\chi>\chi_2$), showing the above
result explicitly.  Together with the observation that ${\cal F}$
has its maximum at $\theta=0$ for given values of $m_{vis}^{(i)}$
and $m_\chi$, the above result assures that the global maximum of
the $m_{T2}$ function ${\cal F}$ over the 3-dimensional event space
parameterized by $\{m_{vis}^{(i)},\theta\}$ is given by
\begin{eqnarray}
{\cal F}^{\rm max}(m_{\chi})=\left\{\begin{array}{ll} {\cal F}^{\rm
max}_<
  \quad \mbox{for} \,\,
m_{\chi}<m_{\tilde{\chi}_1^0} \\ {\cal F}^{\rm max}_{>}\quad
\mbox{for} \,\, m_{\chi}>m_{\tilde{\chi}_1^0},
\end{array}\right. \end{eqnarray}
where \begin{eqnarray}
 {\cal F}^{\rm max}_< &=&{\cal
F}(m_{vis}^{(1)}=m_{vis}^{\rm min}, m_{vis}^{(2)}=m_{vis}^{\rm
min},\theta=0,m_\chi),\nonumber \\{\cal F}^{\rm max}_{>} &=&{\cal
F}(m_{vis}^{(1)}=m_{vis}^{\rm max}, m_{vis}^{(2)}=m_{vis}^{\rm
max},\theta=0,m_\chi).
\end{eqnarray}

\end{document}